# The Science Performance of JWST as Characterized in Commissioning

Jane Rigby[1,46], Marshall Perrin[2], Michael McElwain[1], Randy Kimble[1], Scott Friedman[2], Matt Lallo[2], René Doyon[3], Lee Feinberg[1], Pierre Ferruit[4], Alistair Glasse[5], Marcia Rieke[6], George Rieke[6], Gillian Wright[5], Chris Willott[7], Knicole Colon[1], Stefanie Milam[1], Susan Neff[1], Christopher Stark[1], Jeff Valenti[2], Jim Abell[1], Faith Abney[2], Yasin Abul-Huda[2], D. Scott Acton[8], Evan Adams[2], David Adler[2], Jonathan Aguilar[2], Nasif Ahmed[2], Loïc Albert[3], Stacey Alberts[6], David Aldridge[2], Marsha Allen[2], Martin Altenburg[10], Javier Álvarez-Márquez[11], Catarina Alves de Oliveira[4], Greg Andersen[1], Harry Anderson[2], Sara Anderson[2], Ioannis Argyriou[12], Amber Armstrong[2], Santiago Arribas[11], Etienne Artigau[3], Amanda Arvai[2], Charles Atkinson[13], Gregory Bacon[2], Thomas Bair[2], Kimberly Banks[1], Jaclyn Barrientes[2], Bruce Barringer[2], Peter Bartosik[9], William Bast[2], Pierre Baudoz[14], Thomas Beatty[6], Katie Bechtold[2], Tracy Beck[2], Eddie Bergeron[2], Matthew Bergkoetter[1], Rachana Bhatawdekar[4], Stephan Birkmann[4], Ronald Blazek[2], Claire Blome[2], Anthony Boccaletti[14], Torsten Böker[4], John Boia[2], Nina Bonaventura[15], Nicholas Bond[1], Kari Bosley[2], Ray Boucarut[1], Matthew Bourque[2], Jeroen Bouwman[16], Gary Bower[2], Charles Bowers[1], Martha Boyer[2], Larry Bradley[2], Greg Brady[2], Hannah Braun[2], David Breda[17], Pamela Bresnahan[2], Stacey Bright[2], Christopher Britt[2], Asa Bromenschenkel[2], Brian Brooks[2], Keira Brooks[2], Bob Brown[8], Matthew Brown[2], Patricia Brown[2], Andy Bunker[18], Matthew Burger[2], Howard Bushouse[2], Steven Cale[1], Alex Cameron[18], Peter Cameron[9], Alicia Canipe[2], James Caplinger[2], Francis Caputo[2], Mihai Cara[2], Larkin Carey[8], Stefano Carniani[19], Maria Carrasquilla[8], Margaret Carruthers[2], Michael Case[2], Riggs Catherine[2], Don Chance[2], George Chapman[2], Stéphane Charlot[20], Brian Charlow[2], Pierre Chayer[2], Bin Chen[2], Brian Cherinka[2], Sarah Chichester[2], Zack Chilton[2], Taylor Chonis[8], Mark Clampin[1], Charles Clark[1], Kerry Clark[2], Dan Coe[2], Benee Coleman[2], Brian Comber[1], Tom Comeau[2], Dennis Connolly[1], James Cooper[1], Rachel Cooper[2], Eric Coppock[8], Matteo Correnti[2], Christophe Cossou[21], Alain Coulais[14], Laura Coyle[8], Misty Cracraft[2], Mirko Curti[22], Steven Cuturic[9], Katherine Davis[2], Michael Davis[1], Bruce Dean[1], Amy DeLisa[1], Wim deMeester[12], Nadia Dencheva[2], Nadezhda Dencheva[2], Joseph DePasquale[2], Jeremy Deschenes[2], Örs Hunor Detre[16], Rosa Diaz[2], Dan Dicken[5], Audrey DiFelice[2], Matthew Dillman[2], William Dixon[2], Jesse Doggett[2], Tom Donaldson[2], Rob Douglas[2], Kimberly DuPrie[2], Jean Dupuis[23], John Durning[1], Nilufar Easmin[2], Weston Eck[2], Chinwe Edeani[2], Eiichi Egami[6], Ralf Ehrenwinkler[10], Jonathan Eisenhamer[2], Michael Eisenhower[24], Michelle Elie[2], James Elliott[2], Kyle Elliott[2], Tracy Ellis[2], Michael Engesser[2], Nestor Espinoza[2], Odessa Etienne[2], Mireya Etxaluze[25], Patrick Falini[2], Matthew Feeney[2], Malcolm Ferry[26], Joseph Filippazzo[2], Brian Fincham[2], Mees Fix[2], Nicolas Flagey[2], Michael Florian[6], Jim Flynn[13], Erin Fontanella[2], Terrance Ford[1], Peter Forshay[2], Ori Fox[2], David Franz[1], Henry Fu[13], Alexander Fullerton[2], Sergey Galkin[2], Anthony Galyer[1], Macarena García Marín[4], Jonathan P. Gardner[1], Lisa Gardner[2], Dennis Garland[2], Bruce Garrett[2], Danny Gasman[12], Andras Gaspar[6], Daniel Gaudreau[23], Peter Gauthier[2], Vincent Geers[5], Paul Geithner[1], Mario Gennaro[2], Giovanna Giardino[27], Julien Girard[2], Mark Giuliano[2], Kirk Glassmire[2], Adrian Glauser[28], Stuart Glazer[1], John Godfrey[2], David Golimowski[2], David Gollnitz[2], Fan Gong[9], Shireen Gonzaga[2], Michael Gordon[8], Karl Gordon[2], Paul Goudfrooij[2], Thomas Greene[29], Matthew Greenhouse[1], Stefano Grimaldi[8], Andrew Groebner[2], Timothy Grundy[25], Pierre Guillard[20], Irvin Gutman[2], Kong Q. Ha[1], Peter Haderlein[17], Andria Hagedorn[13], Kevin Hainline[6], Craig Haley[9], Maryam Hami[2], Forrest Hamilton[2], Heidi Hammel[30], Carl Hansen[2], Tom Harkins[13], Michael Harr[2], Jessica Hart[2], Quyen Hart[2], George Hartig[2], Ryan Hashimoto[13], Sujee Haskins[1], William Hathaway[2], Keith Havey[1], Brian Hayden[2], Karen Hecht[2], Chris Heller-Boyer[2], Caroline Henriques[2], Alaina Henry[2], Karl Hermann[10], Scarlin Hernandez[2], Brigette Hesman[2], Brian Hicks[8], Bryan Hilbert[2], Dean Hines[2], Melissa Hoffman[2], Sherie Holfeltz[2], Bryan J. Holler[2], Jennifer Hoppa[2], Kyle Hott[13], Joseph M. Howard[1], Rick Howard[31], Alexander Hunter[2], David Hunter[2], Brendan Hurst[2], Bernd Husemann[16], Leah Hustak[2], Luminita Ilinca Ignat[23], Garth Illingworth[32], Sandra Irish[1], Wallace Jackson[13], Amir Jahromi[1], Peter Jakobsen[33], LeAndrea James[2], Bryan James[1], William Januszewski[2], Ann Jenkins[2], Hussein Jirdeh[2], Phillip Johnson[2], Timothy Johnson[2], Vicki Jones[2], Ron Jones[1], Danny Jones[2], Olivia Jones[5], Ian Jordan[2], Margaret Jordan[2], Sarah Jurczyk[8], Alden Jurling[1], Catherine Kaleida[2], Phillip Kalmanson[2], Jens Kammerer[2], Huijo Kang[2], Shaw-Hong Kao[2], Diane Karakla[2], Patrick Kavanagh[34], Doug Kelly[6], Sarah Kendrew[4], Herbert Kennedy[2], Deborah Kenny[2], Ritva Keski-kuha[1], Charles Keyes[2], Richard Kidwell[2], Wayne Kinzel[2], Jeff Kirk[1], Mark Kirkpatrick[13], Danielle Kirshenblat[2], Pamela Klaassen[5], Bryan Knapp[2], J. Scott Knight[8], Perry Knollenberg[13], Robert Koehler[2], Anton Koekemoer[2], Aiden Kovacs[2], Trey Kulp[2], Nimisha Kumari[35], Mark Kyprianou[2], Stephanie La Massa[2], Aurora Labador[2], Alvaro Labiano[11,36], Pierre-Olivier Lagage[21], Charles-Philippe Lajoie[2], Matthew Lallo[2], May Lam[2], Tracy Lamb[2], Scott Lambros[1], Richard Lampenfield[2], James Langston[2], Kirsten Larson[35], David Law[2], Jon Lawrence[1], David Lee[5], Jarron Leisenring[6], Kelly Lepo[2], Michael Leveille[2], Nancy Levenson[2], Marie Levine[17], Zena Levy[2], Dan Lewis[26], Hannah Lewis[2], Mattia Libralato[35], Paul Lightsey[8], Miranda Link[2], Lily Liu[2], Amy Lo[13], Alexandra Lockwood[2], Ryan Logue[2], Chris Long[2], Douglas Long[2], Charles Loomis[2], Marcos Lopez-Caniego[37], Jose Lorenzo Alvarez[4], Jennifer Love-Pruitt[13], Adrian Lucy[2],

Nora Luetzgendorf[4], Peiman Maghami[1], Roberto Maiolino[22], Melissa Major[2], Sunita Malla[2], Eliot Malumuth[1], Elena Manjavacas[35], Crystal Mannfolk[2], Amanda Marrione[2], Anthony Marston[4], André Martel[2], Marc Maschmann[10], Gregory Masci[2], Michaela Masciarelli[8], Michael Maszkiewicz[23], John Mather[1], Kenny McKenzie[1], Brian McLean[2], Matthew McMaster[2], Katie Melbourne[8], Marcio Meléndez[2], Michael Menzel[1], Kaiya Merz[2], Michele Meyett[2], Luis Meza[13], Cherie Miskey[1], Karl Misselt[6], Christopher Moller[1], Jane Morrison[6], Ernie Morse[2], Harvey Moseley[38], Gary Mosier[1], Matt Mountain[30], Julio Mueckay[8], Michael Mueller[39], Susan Mullally[2], Jess Murphy[8], Katherine Murray[2], Claire Murray[2], David Mustelier[40], James Muzerolle[2], Matthew Mycroft[17], Richard Myers[13], Kaila Myrick[2], Shashvat Nanavati[13], Elizabeth Nance[2], Omnarayani Nayak[2], Bret Naylor[17], Edmund Nelan[2], Bryony Nickson[2], Alethea Nielson[1], Maria Nieto-Santisteban[2], Nikolay Nikolov[2], Alberto Noriega-Crespo[2], Brian O'Shaughnessy[2], Brian O'Sullivan[4], William Ochs[1], Patrick Ogle[2], Brenda Oleszczuk[2], Joseph Olmsted[2], Shannon Osborne[2], Richard Ottens[1], Beverly Owens[2], Camilla Pacifici[2], Alyssa Pagan[2], James Page[2], Sang Park[24], Keith Parrish[1], Polychronis Patapis[28], Lee Paul[2], Tyler Pauly[2], Cheryl Pavlovsky[2], Andrew Pedder[2], Matthew Peek[2], Maria Pena-Guerrero[2], Konstantin Penanen[17], Yesenia Perez[2], Michele Perna[11], Beth Perriello[2], Kevin Phillips[1], Martin Pietraszkiewicz[13], Jean-Paul Pinaud[13], Norbert Pirzkal[35], Joseph Pitman[41], Aidan Piwowar[9], Vera Platais[2], Danielle Player[2], Rachel Plesha[2], Joe Pollizi[2], Ethan Polster[2], Klaus Pontoppidan[2], Blair Porterfield[2], Charles Proffitt[2], Laurent Pueyo[2], Christine Pulliam[2], Brian Quirt[2], Irma Quispe Neira[2], Rafael Ramos Alarcon[2], Leah Ramsay[2], Greg Rapp[8], Robert Rapp[1], Bernard Rauscher[1], Swara Ravindranath[2], Timothy Rawle[4], Michael Regan[2], Timothy A. Reichard[1], Carl Reis[42], Michael E. Ressler[17], Armin Rest[2], Paul Reynolds[13], Timothy Rhue[2], Karen Richon[1], Emily Rickman[4], Michael Ridgaway[2], Christine Ritchie[2], Hans-Walter Rix[16], Massimo Robberto[2], Gregory Robinson[31], Michael Robinson[2], Orion Robinson[2], Frank Rock[2], David Rodriguez[2], Bruno Rodriguez Del Pino[11], Thomas Roellig[29], Scott Rohrbach[1], Anthony Roman[2], Fred Romelfanger[2], Perry Rose[2], Anthony Roteliuk[13], Marc Roth[13], Braden Rothwell[2], Neil Rowlands[9], Arpita Roy[2], Pierre Royer[12], Patricia Royle[2], Chunlei Rui[13], Peter Rumler[4], Joel Runnels[8], Melissa Russ[2], Zafar Rustamkulov[43], Grant Ryden[13], Holly Ryer[2], Modhumita Sabata[2], Derek Sabatke[8], Elena Sabbi[2], Bridget Samuelson[13], Benjamin Sapp[2], Bradley Sappington[2], B. Sargent[2,43], Arne Sauer[10], Silvia Scheithauer[16], Everett Schlawin[6], Joseph Schlitz[2], Tyler Schmitz[2], Analyn Schneider[17], Jürgen Schreiber[16], Vonessa Schulze[2], Ryan Schwab[2], John Scott[2], Kenneth Sembach[2], Clare Shanahan[2], Bryan Shaughnessy[25], Richard Shaw[2], Nanci Shawger[13], Christopher Shay[2], Evan Sheehan[1], Sharon Shen[2], Allan Sherman[1], Bernard Shiao[2], Hsin-Yi Shih[2], Irene Shivaei[6], Matthew Sienkiewicz[2], David Sing[43], Marco Sirianni[4], Anand Sivaramakrishnan[2], Joy Skipper[2], G. C. Sloan[2], Christine Slocum[2], Steven Slowinski[2], Erin Smith[1], Eric Smith[31], Denise Smith[2], Corbett Smith[1], Gregory Snyder[2], Warren Soh[9], Sangmo Tony Sohn[2], Christian Soto[2], Richard Spencer[2], Scott Stallcup[2], John Stansberry[2], Carl Starr[1], Elysia Starr[13], Alphonso Stewart[1], Massimo Stiavelli[2], Amber Straughn[1], David Strickland[2], Jeff Stys[2], Francis Summers[2], Fengwu Sun[6], Ben Sunnquist[2], Daryl Swade[2], Michael Swam[2], Robert Swaters[2], Robby Swoish[13], Joanna M. Taylor[2], Rolanda Taylor[2], Maurice Te Plate[4], Mason Tea[2], Kelly Teague[2], Randal Telfer[2], Tea Temim[44], Deepashri Thatte[2], Christopher Thompson[2], Linda Thompson[2], Shaun Thomson[1], Tuomo Tikkanen[45], William Tippet[2], Connor Todd[2], Sharon Toolan[2], Hien Tran[2], Edwin Trejo[2], Justin Truong[2], Chris Tsukamoto[13], Samuel Tustain[25], Harrison Tyra[2], Leonardo Ubeda[2], Kelli Underwood[2], Michael Uzzo[2], Julie Van Campen[1], Thomas Vandal[3], Bart Vandenbussche[12], Begoña Vila[1], Kevin Volk[2], Glenn Wahlgren[2], Mark Waldman[1], Chanda Walker[8], Michel Wander[23], Christine Warfield[2], Gerald Warner[9], Matthew Wasiak[1], Mitchell Watkins[2], Andrew Weaver[1], Mark Weilert[17], Nick Weiser[8], Ben Weiss[13], Sarah Weissman[2], Alan Welty[2], Garrett West[8], Lauren Wheate[1], Elizabeth Wheatley[2], Thomas Wheeler[2], Rick White[2], Kevin Whiteaker[2], Paul Whitehouse[1], Jennifer Whiteleather[2], William Whitman[2], Christina Williams[6], Christopher Willmer[6], Scott Willoughby[13], Andrew Wilson[9], Gregory Wirth[8], Emily Wislowski[2], Erin Wolf[8], David Wolfe[2], Schuyler Wolff[6], Bill Workman[2], Ray Wright[8], Carl Wu[1], Rai Wu[2], Kristen Wymer[2], Kayla Yates[2], Christopher Yeager[2], Jared Yeates[9], Ethan Yerger[2], Jinmi Yoon[2], Alice Young[4], Susan Yu[2], Dean Zak[2], Peter Zeidler[35], Julia Zhou[9], Thomas Zielinski[1], Cristian Zincke[1], and Stephanie Zonak[2]

[1] NASA's Goddard Space Flight Center, USA; Jane.Rigby@nasa.gov
[2] Space Telescope Science Institute, USA
[3] Université de Montréal, Canada
[4] European Space Agency
[5] UK Astronomy Technology Centre, UK
[6] University of Arizona, USA
[7] National Research Council Canada (Herzberg), Canada
[8] Ball Aerospace, USA
[9] Honeywell International, USA
[10] Airbus Defence and Space GmbH, Germany
[11] Centro de Astrobiología (INTA/CSIC), Spain
[12] Katholieke Universiteit Leuven, Belgium
[13] Northrop Grumman Space Systems, USA
[14] Observatoire de Paris, France
[15] Niels Bohr Institute, Denmark
[16] Max-Planck-Institut für Astronomie, Germany

[17] Jet Propulsion Laboratory, California Institute of Technology, USA
[18] University of Oxford, UK
[19] Scuola Normale Superiore di Pisa, Italy
[20] Institut d'Astrophysique de Paris, France
[21] Commissariat à l'énergie atomique et aux énergies alternatives, France
[22] University of Cambridge, UK
[23] Canadian Space Agency, Canada
[24] Center for Astrophysics | Harvard & Smithsonian, USA
[25] Rutherford Appleton Laboratory, UK
[26] Lockheed Martin Advanced Technology Center, USA
[27] ATG Europe for the European Space Agency, The Netherlands
[28] Eidgenössische Technische Hochschule, Switzerland
[29] NASA's Ames Research Center, USA
[30] Association of Universities for Research in Astronomy, USA
[31] National Aeronautics and Space Administration, USA
[32] University of California Santa Cruz, USA
[33] University of Copenhagen, Denmark
[34] Dublin Institute for Advanced Studies, Ireland
[35] AURA for the European Space Agency, USA
[36] Telespazio UK for the European Space Agency, UK
[37] Aurora Technology for the European Space Agency, The Netherlands
[38] Quantum Circuits, USA
[39] University of Groningen, The Netherlands
[40] ATA Aerospace, USA
[41] Heliospace, USA
[42] NASA's Johnson Space Center, USA
[43] John Hopkins University, USA
[44] Princeton University, USA
[45] University of Leicester, UK

## Abstract

This paper characterizes the actual science performance of the James Webb Space Telescope (JWST), as determined from the six month commissioning period. We summarize the performance of the spacecraft, telescope, science instruments, and ground system, with an emphasis on differences from pre-launch expectations. Commissioning has made clear that JWST is fully capable of achieving the discoveries for which it was built. Moreover, almost across the board, the science performance of JWST is better than expected; in most cases, JWST will go deeper faster than expected. The telescope and instrument suite have demonstrated the sensitivity, stability, image quality, and spectral range that are necessary to transform our understanding of the cosmos through observations spanning from near-earth asteroids to the most distant galaxies.

*Key words:* Observatories – Infrared astronomy – Astronomical instrumentation

## 1. Introduction

This paper characterizes the delivered science performance of JWST, as determined through the six month commissioning period that ended on 2022 July 12, and describes how the actual performance differs from pre-launch expectations.

JWST is a large (6.6 m), cold (<50 K), infrared-optimized observatory (Gardner et al. 2022, this issue) with a segmented mirror design, that was launched on 2021 December 25 and is now in science operations. The project is an international collaboration among NASA, the European Space Agency (ESA), and the Canadian Space Agency (CSA). The review article Gardner et al. (2006) described JWST's design and science goals, which are divided into four themes: "End of the Dark Ages: First Light and Reionization;" "Assembly of Galaxies;" "Birth of Stars and Protoplanetary Systems;" and "Planetary Systems and the Origins of Life." The design and architecture of JWST are described in Menzel et al. (2022) (this issue).

During the six-month commissioning period of JWST, the mission team worked with dedication and focus to prepare the observatory for science operations. A key part of commissioning activities was characterizing the on-orbit performance of the observatory, including the performance of the spacecraft, telescope, science instruments, and ground system. This paper summarizes those results, as drawn from many activities and analyses from the commissioning period.

[46] Author to whom any correspondence should be addressed.

The expected pre-launch performance was incorporated into the pre-launch versions of the exposure time calculator (ETC)[47] and the JWST backgrounds tool.[48] These tools have now been revised to reflect actual performance, to support the Cycle 2 Call for Proposals.

To inform the scientific community, an early version of this paper was posted online (arXiv:2207.05632) at the end of commissioning on 2022 July 12. For the PASP special issue on JWST, some description of science instrument performance has been moved to companion papers: Böker et al. (2023) for NIRSpec; Doyon et al. (2022) for NIRISS/FGS, Rieke et al. (2023) for NIRCam, and Wright et al. (2023) for MIRI. Some description of the JWST background levels has been moved to a different PASP paper (Rigby et al. 2023). The companion PASP paper by McElwain et al. (2023) describes more fully the telescope element of JWST.

The transformative scientific performance of JWST is the result of the collective effort, spanning decades, of thousands of individuals from multiple institutions. The authors acknowledge the tremendous amount of work by the entire international team to bring JWST through commissioning into science operations.

## 2. Spacecraft

### 2.1. Orbit

The Ariane 5 rocket that launched JWST on 2021 December 25 UT injected it into an orbit that was well within specification, with a semimajor axis approximately $0.5\sigma$ larger than nominal. This very slightly "hot" injection state had a semimajor axis of 542,120.1 km versus the nominal of 536,533.8 km, and a delta-v at launch vehicle separation within $3\ \mathrm{m\ s^{-1}}$ of the target velocity of $10,089\ \mathrm{m\ s^{-1}}$. This accurate launch, in combination with three on-time, nominal mid-course corrections, minimized propellant consumption and delivered JWST to a nominal orbit around the second Earth-Sun Lagrange point (known as L2). This orbit fully complies with all geometry requirements, and supports communications with the Science & Operations Center using the Deep Space Network.

Orbit around L2 is maintained through regular station-keeping burns, which are scheduled every three weeks. As of 2022 July 12, there have been four station-keeping burns, with typical durations of tens of seconds. During commissioning, three station-keeping burns were skipped because the computed correction was negligibly small.

### 2.2. Predicted Lifetime of Consumables

There are no consumable cryogens onboard JWST; the telescope and the science instruments are passively cooled by the sunshield and radiators, and MIRI's active cryocooler recycles its helium. The only onboard consumables are propellant: fuel and oxidizer. Before launch, JWST was required to carry propellant for at least 10.5 yr of mission lifetime. Now that JWST is in orbit around L2, it is clear that the remaining propellant will last for more than 20 yr of mission lifetime. This fortunate surfeit has multiple causes: an accurate launch; launch on a day that required relatively less energy to get to L2 than most other possible launch dates; three timely and accurate mid-course corrections that sent JWST to L2 with the minimum possible propellant usage; and finally, careful stewardship of mass margins by the engineering team over the years, such that the remaining mass margin was used to add more propellant than required, until the tanks were full.

For the remainder of the mission, propellant will be used for two purposes: stationkeeping burns (using fuel and oxidizer) to maintain the orbit around L2, and momentum dumps (using only fuel) to remove momentum from the reaction wheels. Momentum accumulates as solar photons hit the sunshield and impart a net torque, which the reaction wheels resist by spinning up. The actual rate at which the observatory builds up momentum is within specifications and is well below worst-case allocations, which further contributes to propellant lifetime. While the detailed propellant usage depends on orientation, which is set by the observing schedule, the big picture is that JWST has sufficient propellant onboard to support science operations for more than 20 yr.

### 2.3. Projected Observatory Lifetime

At this point, it is not clear what will determine the duration of JWST's mission. The mirrors and sunshield are expected to slowly degrade from micrometeoroid impacts; the detectors are expected to experience cumulative slow damage from charged particles; the sunshield and multilayer insulation will degrade from space weathering; the spacecraft was designed for a five year mission (as is standard for NASA science missions); and the science instruments include many moving parts at cryogenic temperatures. These sources of degradation were all taken into account in the design of JWST, with performance margins set so that JWST will still perform after many years of operation. At present, the largest source of uncertainty is long term effects of micrometeoroid impacts that slowly degrade the primary mirror. As discussed in Section 4.7, the single micrometeorite impact that occurred between 2022 May 22 and 24 UT exceeded prelaunch expectations of damage for a single micrometeoroid,[49] triggering further investigation and modeling by the JWST Project. The

[47] https://jwst.etc.stsci.edu/

[48] https://jwst-docs.stsci.edu/jwst-other-tools/jwst-backgrounds-tool

[49] Pre-launch projections, informed by micrometeoroid population models and experimental studies and numerical simulations of impacts to beryllium mirrors, predicted that on average each segment would receive a cumulative total of 16 nm added WFE over six years. The May impact resulted in one segment receiving more than 10 times that average in a single event.

Project is actively working this issue to ensure a long, productive science mission with JWST.

### 2.4. Slew Speed and Settle Times

The slew speed and settle time are important drivers for efficient operations as the observatory changes pointing to look at different targets on sky and carry out orbital stationkeeping and momentum unloading maneuvers. Slews use all six reaction wheels on the spacecraft to repoint the observatory, although it is possible to control with five reaction wheels. The control system was designed to slew 90° in less than 60 minutes, which has been demonstrated during commissioning.

Achieved slew speeds and durations to repoint to new targets at the start of each visit are broadly consistent with pre-flight expectations, such as the timing model encoded in the Astronomer's Proposal Tool (APT), plus typically 2 minutes duration for control overheads and settling.

At the end of a slew, pointing transients are observed while the observatory settles. The pointing settles in ~30 s with damping from isolators between the telescope and spacecraft bus. Re-pointing maneuvers can generate fuel slosh, which is at ~0.045 Hz and not compensated by the fine guidance control system. The slew rate profile has been tuned to reduce the excitation of the fuel slosh mode. Measurements of line-of-sight pointing performance (Section 3.3 below) confirm that the resulting effect of fuel slosh on pointing is <0.5 milliarcseconds (mas).

After slewing to a new target, the pointing stabilizes quickly in less time than it takes the FGS guide star acquisition process to complete. This was not initially the case; in the first months of commissioning, long slew settling times and high image motion impacted many observations, and required efforts to investigate, diagnose, and mitigate. Adjustments to Attitude Control System parameters and several software patches dramatically improved slew settling performance and resolved this issue. Users examining very early commissioning data (prior to mid 2022 April), particularly images taken in coarse point mode without fine guiding, should be aware of this caveat.

### 2.5. Thermal

The cooldown of the telescope and science instruments was nominal and closely matched predictions. As predicted, the primary mirror segments have cooled to temperatures of 35–55 K, with the hotter segments those closest to the sunshield. The secondary mirror has cooled to 29.3 K, the near-IR instruments to 35–39 K, and MIRI to 6.4 K. The MIRI cooler achieves this temperature at nominal, pre-flight predicted performance levels and has no perceptible effect on pointing stability (e.g., jitter) or on the performance of the other instruments. Since cooling to operational temperatures, these temperatures have remained extremely stable with time. For the telescope mirrors and structures, the stability is within the 40 mK noise of the temperature sensors on those components. The science instrument temperatures vary based on their activity, but they are also within the 10 mK noise for the instrument sensors. Any resulting temperature change impact can only be identified optically (see Section 4.5). No long term temperature drift has been detected since achieving final cooldown conditions. Temperatures of detector focal plane arrays are actively stabilized to a precision of a few mK.

The observatory was designed to minimize ice deposition on the optical surfaces. Sensitive components of the science instruments and the telescope's fine steering mirror were heated through the cooldown to prevent ice accumulation. Throughput measurements detect no spectral signatures of ice, which constrains any ice deposition that may have occurred to be far better (less ice) than the requirements.

Components within the spacecraft bus (computer, communications, cryocooler compressors and electronics, attitude control and propulsion systems, etc.) are comfortably within required operating temperature ranges regardless of pointing direction or telescope activities. Instrument electronics housed on the cold side of the Observatory are also within required operating range and are under tight temperature control to minimize any temperature-induced distortions (see Section 4.5.3). All heaters on the observatory are functional on prime circuits and demonstrating expected margins.

### 2.6. Sunshield Performance

The shape of the deployed sunshield affects the temperature and thermal stability of JWST, the amount of scattered light from the Earth, Moon, and stars, and the background levels at the longer wavelengths. Telemetry (microswitch, motor, thermal, power, and inertial reference unit) indicated a successful deployment ending with a nominal deployed shape. There are no subsequent indications of any issue with the deployed shape, from the many thermal sensors onboard or from the background levels seen in the science instruments.

The shape of the deployed sunshield affects how quickly the observatory builds up angular momentum from solar photons, which is then stored in the reaction wheels and must be dumped periodically using thrusters. The sunshield geometry was designed to minimize momentum accumulation. The measured torque table on-orbit is consistent with the pre-launch model within the allocated uncertainties. As noted above, this implies a lower rate of fuel use for momentum management.

### 2.7. Other Spacecraft Performance

After optimization of the solar array regulator settings, JWST is now generating 1.5 kW to match the power load, with a capability of >2 kW—as such, the power margins are comfortable. JWST is projected to have a tight (11%) margin on data downlink during Cycle 1. The project is working closely with NASA's Deep Space Network (DSN) to resolve all issues and ensure DSN can adequately support all Cycle 1 science and beyond. The JWST spacecraft has all the redundancy it had at launch. Of the 344 single point failures[50] that were present at launch—almost all of them related to deployments—only 49 remain; these are common to most science missions (for example, only one set of propellant tanks, only one high gain antenna). Fifteen of the remaining single point failure items are associated with the science instruments —any future failure with these items would degrade science performance but would not end the mission.

### 2.8. Fault Management

JWST has a robust fault management system that makes use of its redundant components to ensure the observatory remains safe. With any new spacecraft/observatory, the first couple months after launch provide the operations team an opportunity to learn how the vehicle performs on-orbit and adjust the fault management and system parameters to fine tune the system behavior. During commissioning, JWST experienced seven safe mode entries (six safe haven and one inertial point mode). Early science operations (2022 July - December) experienced one safe haven entry and four entries into inertial point mode. The majority of these safe mode entries were a result of on-orbit learning of the nuances of the control system behavior and system level interactions. All of the safe mode entries' underlying causes are understood and several updates to the fault management and control system set items have been loaded to flight software to prevent the issues from recurring.

Likewise, flight software has placed individual instruments into safe modes on several occasions in responses to unexpected telemetry values or conflicting commands. In all cases the underlying issues were quickly understood, and appropriate steps were taken to bring the instrument back into operation and prevent recurrence of the fault. JWST's onboard event-driven operations system works as intended to allow the observatory to flexibly continue observations when an observation cannot execute due to an instrument fault or a guide star acquisition error; in such cases, JWST will automatically move on to the next observation in the onboard observation plan.

## 3. Pointing and Guiding

Observatory attitude control and line-of-sight stabilization to achieve JWST's stringent requirements present a complex engineering challenge, involving a sophisticated interplay between hardware systems across the entire observatory and arguably the most complex set of flight software components. Line-of-sight pointing control with JWST requires the interconnected operation of star trackers, inertial reference units (gyroscopes), fine Sun sensors, reaction wheels, the telescope's fine steering mirror, the fine guidance sensor, various target acquisition modes in science instruments, flight software in both the spacecraft and science instrument computers, and ground software systems for guide star selection and observation planning, as well as the structural dynamics of the deployed observatory.

It is inherently not possible to test those systems together in an end-to-end fashion on the ground. As such, on-orbit performance offers the first chance to characterize the system as a whole. As commissioning concludes, the pointing performance of the observatory meets or exceeds expectations.

### 3.1. Pointing Accuracy after Guide Star Acquisition

Absolute pointing accuracy after guide star acquisition is excellent. Observed pointing offsets are generally below $0\farcs19$ ($1\sigma$, radial). When systematic offsets are removed by improved astrometric calibration between the Guiders and the other science instruments (SIs), the residual scatter around the desired target position is generally below $0\farcs10$ ($1\sigma$, radial) , better than the prelaunch predicted value of $0\farcs14$, which is in turn significantly better than the required value of $1\farcs0$. Updates for such systematic offsets are in progress. This confirms expectations that many integral field spectroscopy (IFS) observations may omit target acquisition and rely solely on guide star acquisition to achieve the necessary science pointing. The excellent pointing performance can be credited to JWST's own systems, as well as the high accuracy and precision of the guide star catalog enabled by the Gaia mission.

Early in commissioning, some observations had significantly larger pointing offsets, of order $\sim 1\farcs5$, due to catalog cross-matching issues that occurred when data from Gaia and other catalogs such as 2MASS were combined to produce the current JWST guide star catalog. Updated guidelines for selecting guide/reference stars was implemented 2022 May 5, and non-stars were disallowed as reference objects 2022 June 24; both greatly reduced mis-identification. A future version of the catalog is planned for 2023 which is expected to provide further improvements.

Pointing repeatability is likewise excellent: independent separate observations returning to the same target and at the same position angle generally result in identical target pointings on a detector to within $<0\farcs1$.

### 3.2. Pointing Accuracy After Target Acquisition

For observing scenarios requiring better final pointing accuracy than provided by the guide star catalog alone, the instruments offer onboard Target Acquisition procedures to place targets where desired within a science instrument field, e.g., centered on a coronagraphic spot or null, or in a repeatable position for NIRISS Aperture Masking Interferometry or Single Object Slitless Spectroscopy (SOSS). There are several distinct versions of target acquisition for the various instrument modes. These onboard processes have been confirmed to meet their requirements, typically yielding target positions within a few mas rms per axis in the near-IR instruments, and slightly larger in MIRI with its broader longer-wavelength PSFs. NIRCam coronagraphy, the final mode to have its Target Acquisition process evaluated, currently experiences larger offsets, which will be alleviated by future calibrations—even so, the Small Grid Dithers recommended for that mode cover the range of uncertainty and yield excellent coronagraphic contrast.

[50] The NASA Goddard GOLD rules, https://standards.nasa.gov/standard/GSFC/GSFC-STD-1000, define a failure in this context as preventing the mission from fully meeting level 1 requirements; this is a stricter definition than failure meaning loss of the mission.

The algorithm for NIRSpec Multi-Object Spectroscopy is the most complex, because it derives a roll correction for the observatory position angle as well as the usual *x, y* positional offsets. That target acquisition required great care in implementation, and was the most challenging type of target acquisition to get working during commissioning; it is now working within requirements when the requisite attention is paid to the accuracy of the input reference star information.

### *3.3. Guiding Precision and Line of Sight Pointing Stability*

The pointing stability of the line of sight under fine guidance control is superb, several times better than requirements. The FGS sensing precision, parameterized as the Noise Equivalent Angle (NEA, an equivalent jitter angle calculated based on centroid precision and S/N) is usually $\sim$1 mas ($1\sigma$ per axis), as compared to the requirement of 4 mas; the NEA is usually symmetric in $x$ and $y$.

The achieved line-of-sight jitter as seen in the science instruments is similarly good, even when measured at higher frequency than the 15.6 Hz cadence of the FGS measurements used in calculating the Guider NEA. Jitter measured via high-frequency sampling (every 2.2 ms) using a small NIRCam subarray has also been $\sim$1 mas ($1\sigma$ per axis). For long observations, drift in observatory roll could generate systematic motion of sources in an SI field, as the FGS can not sense or correct roll with the single guide star used in the Fine Guide control loop. However, measurements during a commissioning thermal characterization test indicated that the roll drifts are extremely small, far below requirements, even after a worst-case hot-to-cold slew, and contribute negligibly to the total pointing error. In comparison, pre-flight predictions for pointing stability were in the range 6–7 mas ($1\sigma$ per axis).

The FGS guide star magnitude scale matches the 2MASS *J*-band, Vega scale, with guide star brightnesses of $12.5 < J < 18$ allowed for fixed targets. Dimmer stars will typically have a larger NEA, due to the lower signal to noise per time step, but with NEA that is still well within the requirement. The guide star selection system preferentially selects the brightest available guide star for a given pointing. In practice the achieved NEA can vary by $\sim$0.3 mas between dithers on the same source. For moving targets, the guide star faint limits are somewhat reduced to 16.5 in Guider 1 and 17.0 in Guider 2. Guiding performance can vary under some circumstances. Reasons that can cause a brighter star to appear to register lower counts than expected (and give an increased NEA) include: unflagged bad pixels in the guider box, (flagged) bad pixels exactly coincident with the guider star's position, guiding on an extended galaxy instead of a star, quantum efficiency variations due to cross hatching on the detector surface, or a guidestar falling at the edge of the centroiding box. Work is in progress to reduce these variations. In most cases, observers should expect excellent stability in their observations.

Very precise flight jitter measurements have yielded detailed insights into observatory dynamics, and confirmed many aspects of preflight integrated modeling. These data show that neither the spacecraft's reaction wheels nor the MIRI cryocooler produce measurable contributions to the line-of-sight jitter. Instead there are two signatures, detectable but at low level, that come from a 0.3 Hz oscillation mode of a vibration damper between the telescope and spacecraft bus, and a $\sim$0.045 Hz oscillation understood to be due to fuel slosh, both consistent with models. The amplitudes of these modes range up to $\sim$0.3 mas each, with variation over time that does not appear correlated with pointing history such as slew distances.

Momentum/reaction wheel operations or unplanned High Gain Antenna motion may occasionally disturb observatory dynamics and degrade pointing stability; these are generally short-lived and fine guiding is expected to be maintained. High Gain Antenna moves are not planned during any science observations except for long duration time series observations.

### *3.4. Precision of Dithering*

JWST has three methods of performing a small repointing, commonly called a dither. Dithers less than 60 mas are executed with the fine steering mirror (FSM) while maintaining closed loop on the guide star. Dithers between 60 mas and 25″ are executed by dropping closed-loop guiding, slewing, and re-entering guiding at the track mode stage. Dithers larger than 25″ are executed by dropping closed-loop guiding, slewing, and re-entering guiding at the guide star acquisition stage.

All three methods of dithers result in an offset precision as measured by the Guider of 1–2 mas. On the sky, the accuracy of the dithers is typically 2–4 mas rms per axis, due to residuals in the Guider's astrometric calibration or small systematic offsets in the Guider's calculated $3 \times 3$ pixel centroids. As reported by the observing SI, offsets for large dithers may differ by another few mas, due to residuals in the astrometric calibration of the SI. As SI astrometric calibrations continue to improve, these residuals may decrease.

**Table 1**
Moving Targets Tested during Commissioning and Early Science Operations

| Moving target (type) | Apparent Rate of Motion ($mas\ s^{-1}$) | Program ID | Instrument/Mode |
|---|---|---|---|
| Jupiter (planet) | 3.3 | 1022 | NIRCam Imaging, NIRISS Imaging, NIRSpec fixed slits and IFS, MIRI MRS and imaging |
| 2516 Roman (MBA) | 4.7 | 1449 | MIRI/Imaging |
| 118 Peitho (MBA) | 4.9 | 1449 | MIRI/ LRS and MRS |
| 6481 Tenzing (MBA) | 5 | 1021 | NIRCam/Imaging |
| 1773 Rumpelstilz (MBA) | 6.6 | 1021 | NIRISS/AMI |
| 216 Kleopatra (MBA) | 11 | 1444 | NIRSpec/ IFS and MOS longslit |
| 2035 Stearns (Mars- crossing asteroid) | 24 | 1021 | NIRCam/Imaging |
| 4015 Wilson-Harrington (Apollo, NEO, PHA) | 40 | 1021 | NIRCam/Imaging |
| 464798 (2004 JX20) (Aten, NEO) | 67 | 1021 | NIRCam/Imaging |
| 411165 (2010 DF1) (Apollo, NEO, PHA) | 90, 110 | 2744 | NIRCam/Imaging |

**Note.** Targets are sorted by apparent rate of motion. The last target on the list was tested during early science operations; the others were tested during commissioning. Target type is listed in parenthesis after the target name, where MBA = main belt asteroid, NEO = near-Earth object, and PHA = potentially hazardous asteroid. Aten and Apollo asteroids have orbits that cross the Earth's orbit around the Sun.

### *3.5. Performance for Tracking Moving Targets*

JWST has a Level 1 requirement to track objects within the solar system at speeds up to 30 mas per second ($mas\ s^{-1}$). In commissioning, tracking was tested at rates from 5 to 67 $mas\ s^{-1}$. These tests verified tracking and science instrument performance for moving targets, including dithering and mosaicking.

All tests of moving targets during commissioning were successful. Centroids showed sub-pixel scatter in all instruments. No test showed elongation of moving targets, as would be expected in the case of poor tracking. rms jitter in the guider was typically $<2$ mas ($1\sigma$, radial), comparable to that seen for fixed targets. Image quality as determined by FWHM measurements of point spread functions was also comparable to that of fixed targets. Table 1 summarizes the moving targets observations during commissioning.

Tracking at faster-than-requirements rates of 30–67 $mas\ s^{-1}$ showed accuracies similar to tracking of slower-moving targets; this potentially opens up science for near-Earth asteroids (NEAs), comets closer to perihelion, and interstellar objects. This capability was pushed further early in science operations, when JWST successfully tested rates up to 110 $mas\ s^{-1}$ ($396'' hr^{-1}$) on near-Earth asteroid 2010 DF1, to observe and track the Double Asteroid Redirection Test (DART) mission target (65803) Didymos at the time when the spacecraft impacted the asteroid moonlet on 2022 September 26, to demonstrate the capability of modifying an object's orbit for the purpose of planetary defense (Thomas, C. et al. in preparation.)

While tracking at super-fast rates has now been demonstrated, the observatory efficiency was poor due to guide star availability and acquisition at these rates. Thus, the new maximum rate of motion being offered is 75 $mas\ s^{-1}$ without limitations, but special permission for rates up to 100 $mas\ s^{-1}$ may be requested and subjected to approval for future cycles.

Observing a bright planet and its satellites and rings was expected to be challenging, due to scattered light that may affect the science instrument employed, but also the fine guidance sensor must track guide stars near the bright planet. Therefore, commissioning included tests of moving target tracking with NIRCam, where Jupiter was incrementally moved from the NIRCam field of view (FOV) to the FGS-2 FOV. See Figure 1. These observations verified the expectation that guide star acquisition works successfully as long as Jupiter is at least $140''$ away from the FGS, consistent with pre-flight modeling.

The other SIs were also tested for efficiency with nearby scattered light, also using the Jupiter system. Preliminary results have measured scattered light contamination on the detectors for all instruments when the planet was not in the primary FOV, which will need to be considered for planning nearby satellite observations. The most notable impact for scattered light is in NIRSpec IFS mode—if a bright planet is on the microshutter array, then light seepage becomes significant.

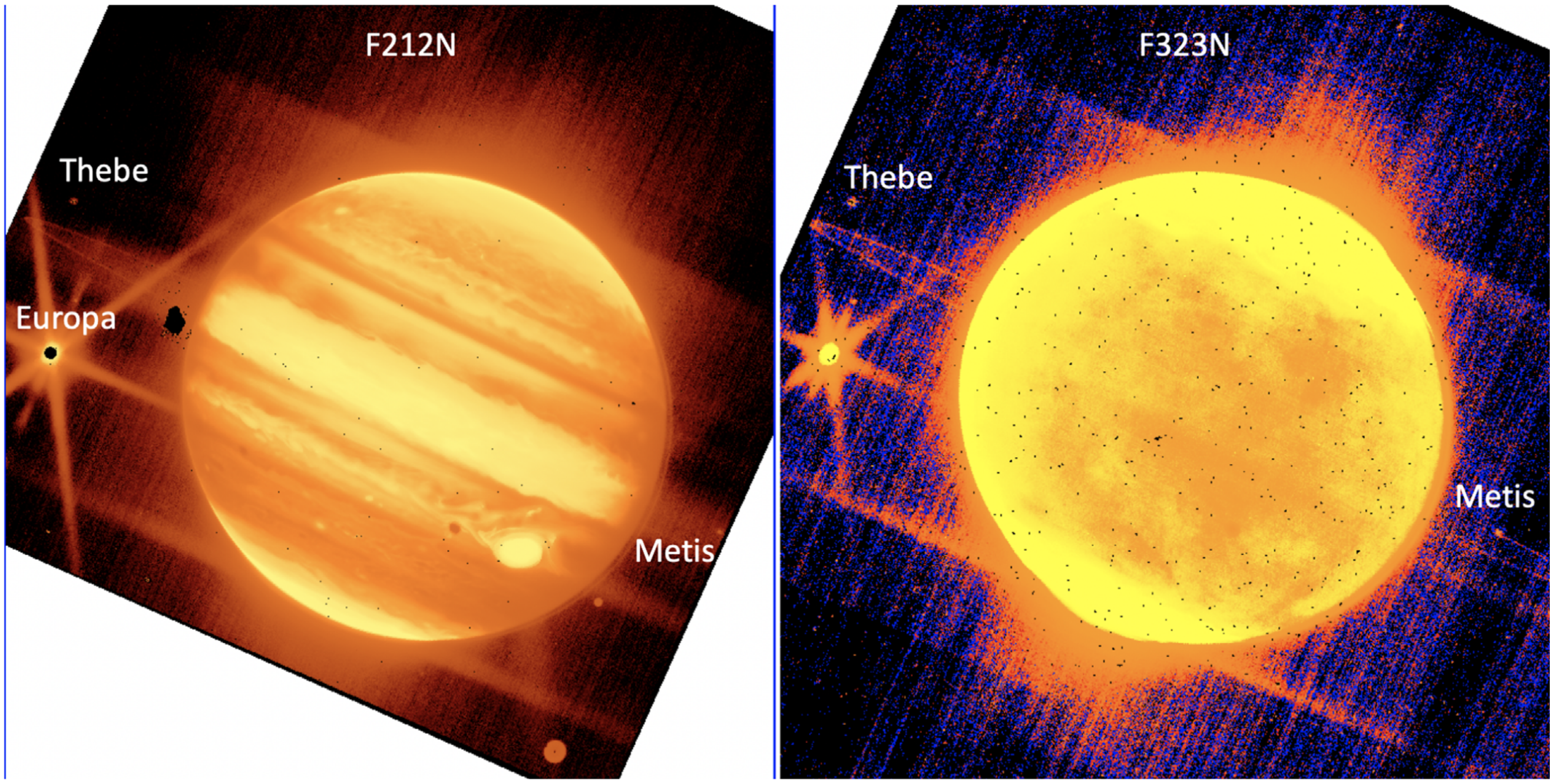

**Figure 1.** NIRCam narrow-band imaging of Jupiter, moons, and ring. PID 1022 demonstrated that JWST can track moving targets even when there is scattered light from a bright Jovian planet. At left is a NIRCam short-wavelength image in filter F212N (2.12 $\mu$m); at right is a NIRCam long-wavelength image in filter F323N (3.23 $\mu$m). The exposure time was 75 s. The Jovian moons Europa, Thebe, and Metis are labeled. The shadow of Europa is also visible, just to the left of the Great Red Spot. The stretch is fairly harsh to bring out the faint moons as well as Jupiter's ring. Each image is from one NIRCam short wavelength detector, which spans 63″.

### *3.6. Success Rate for Closed-loop Guiding*

The closed-loop fine guidance system is one of the most complex aspects of JWST operations, requiring close coordination between several subsystems and multiple control loops running in real time. Activating and tuning the fine guidance system was a significant focus during commissioning. The robustness of the system has steadily improved and continues to do so as reasons for failure are identified and mitigated.

As a snapshot of performance during the later period of commissioning, from May 25 through 2022 June 16, guiding worked successfully ∼93% of the time: guiding was successful on the first try ∼81% of the time, and 12% of the time succeeded on the second or third guide star candidate attempted. (Up to three guide star candidates can be tried in a visit.) In the same time period, guiding failed or skipped 7% of the time.

Several reasons for guiding failure were identified, with steps taken to mitigate them:

1. The "guidestar" is actually a galaxy, but was classified as a star in the guide star catalog. These are flagged in the guide star catalog once found.
2. Attempts to guide on known galaxies frequently failed. Using a known galaxy as a reference source for ID was disallowed starting 2022 June 24.
3. Guide star coordinates, or occasionally brightnesses, in the catalog are incorrect. The catalog is corrected when such errors are found.
4. The guide star catalog contained duplicate entries for some guide stars. This duplication was reduced greatly with new rules for selecting guide stars, implemented 2022 May 5. A new version of the guide star catalog expected in 2023 should largely eliminate this issue.
5. The guide star may be placed on a bad pixel. Bad pixels are flagged once identified.
6. The pointing may not have stabilized sufficiently at the start of the Guide Star ID process.

In Cycle 1, users should expect a closed-loop guiding success rate close to the 93% value that was achieved late in commissioning. That value is, largely by coincidence, very close to the success rate for Hubble guide star acquisitions in recent cycles. Efforts continue to optimize guiding success rates.

## 4. Optical Performance

The image quality achieved by JWST exceeds performance requirements and expectations, having diffraction-limited image quality at wavelengths much lower than requirements, very good stability, and superb throughput. There is not one single factor to credit for the high performance; rather it is the accumulation of performance margins throughout the observatory, and the result of many careful and precise efforts throughout the design, assembly, and alignment of the telescope and instruments.

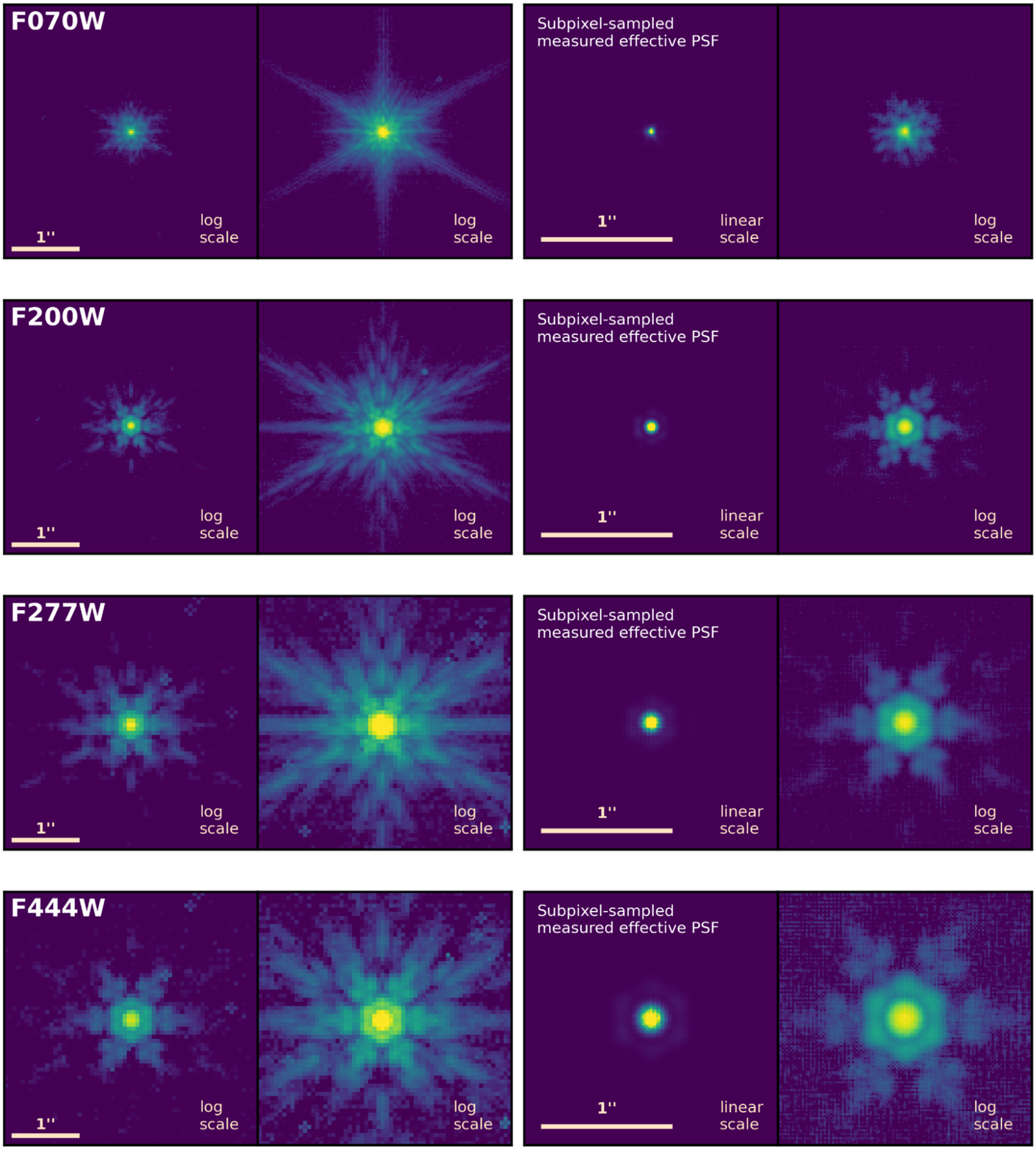

**Figure 2.** Measured Point Spread Functions spanning the full wavelength range of NIRCam.The filters shown are the shortest and longest wide filters in each of the NIRCam short wavelength and long wavelength channels. The left two panels show individual PSFs (single exposure each), on two different logarithmic scales for higher dynamic range. The right two panels show 4× subsampled effective PSFs (ePSFs), generated following the method of Anderson & King (2000) using dithered PSF measurements of many stars; these are shown zoomed in by 2× compared to the left panels. The second column from right, shown on a linear scale, highlights the compact PSF core, while the log display in the other columns emphasizes the diffraction features from the primary mirror geometry. The PSF core is sharp even at the shortest wavelengths in F070W. Data from PIDs 1067 and 1072.

JWST's top-level image quality requirements were defined to achieve Strehl ratio greater than 0.8 at a given wavelength; this can be converted to an equivalent condition of having rms wave front error less than $\lambda/14$, using the optical Marechal approximation (Marechal 1947). These conditions are equivalent; in practice, the wave front error formulation was used for JWST's optical budgets and in-flight wave front sensing measurements.

JWST's requirement was to be diffraction limited at $\lambda = 2\,\mu$m wavelength, which corresponds to 150 nm wave front error by the $\lambda/14$ criterion. The achieved wave front error is routinely between 65 and 70 nm in the NIRCam shortwave channel. During early science operations, wave front control updates have been scheduled when the wave front error exceeds 80 nm (McElwain et al. 2023). Therefore, operationally, the observatory is diffraction limited at $\lambda = 1.1\,\mu$m.

McElwain et al. (2023) (this issue) describes in more detail the telescope deployment and alignment processes as carried out in flight. Here we summarize the achieved optical performance, drawing on extensive telescope wave front sensing and image characterizations throughout commissioning, including a dedicated thermal stability characterization exercise.

### 4.1. Wave front Error and Angular Resolution

The achieved telescope wave front error (WFE), measured at the primary wave front sensing field point in NIRCam module A, is generally in the range 60–80 nanometers rms; it varies on multiple timescales as described below.[51] That WFE contribution sums with field-dependence of the telescope WFE and instrument internal WFE to yield total observatory WFE values which are modestly higher: 75–130 nm depending on instrument, observing mode, and field position. See Table 2. Motions of mirror segments over time can lead occasionally to wave front error levels higher than those values, which are corrected through the routine wave front sensing and control process.

JWST has exquisite image quality across the entire telescope field of view and at all available wavelengths. Expressed relative to wavelength $\lambda$, JWST ranges from $\sim\lambda/10$ for NIRCam F070W to better than $\lambda/100$ for MIRI F1000W and longer. JWST aimed to achieve at $\lambda = 2\,\mu$m a Strehl ratio of 0.8 (corresponding to wave front rms $\lambda/14$ or better), which is considered having diffraction-limited image quality for space-based telescopes. This is achieved with substantial margin, such that the NIRCam short wavelength channel, with typical WFE ∼80 nm, in fact achieves $\lambda/14$ at $\lambda = 1.1\,\mu$m.

**Table 2**
Measured Static Wave front Errors After Multi-instrument Alignment

| Instrument | Measured Observatory Static WFE (nm rms) |
|---|---|
| NIRCam, short-wavelength | 61 ± 8 (module A), 69 ± 11 (module B) |
| NIRCam, long wavelength | 134 ± 38 (module A) 134 ± 39 (module B) |
| NIRISS | 68 ± 12 |
| NIRSpec | 110 ± 20 |
| MIRI | 99 ± 28 |
| FGS | 77 ± 15 (FGS1), 69 ± 8 (FGS2) |

**Note.** Quoted values are the average over multiple field points within each instrument, as measured from multi-instrument multi field sensing during commissioning in 2022 May, and updated for the final focus adjustment at the end of that process. The plus and minus ranges reflect the measured variation in wave front error at different field points within each instrument's field of view. These values represent just the static (non-time-varying) portion of the wavefronts, and include the sum of telescope and instrument WFE together. Additional time-dependent terms sum on top of these at any given time. Data from PID 1465.

That wave front quality yields optical point spread functions (PSFs) with angular resolutions set by the diffraction limit (i.e., PSF full width at half maximum $\sim\lambda/D$) across the full range of available wavelengths. See Figure 2. In particular, even in its shortest filter F070W, the NIRCam short wavelength channel achieves a Strehl ratio of ∼0.6. Though ∼40%–50% of the light is in a diffuse speckle halo at that wavelength, the PSF prior to detector sampling still has a tight core with angular resolution $\sim\lambda/D$. (In that sense, JWST's PSF quality at 0.7 $\mu$m is similar to PSFs achieved at 2 $\mu$m by adaptive optics systems on 8–10 m ground-based telescopes in good conditions.) In practice, angular resolution at the shortest wavelengths (<2 $\mu$m for NIRCam short wavelength, <4 $\mu$m for NIRCam long wavelength or NIRISS) is limited more significantly by detector pixel Nyquist sampling than by optical performance; subpixel dithering and image reconstruction ("drizzling") will be required to make use of the full resolution at these wavelengths.[52]

Wave front sensing confirms the surface quality of the individual mirror segments in space matches closely the mirror surface maps measured during cryogenic testing on the ground. See Figure 3. In other words, after launch into space, and significant thermal contraction and deformation while cooling

[51] The best achieved telescope wavefronts at the completion of alignment were as low as 50 nm rms; the May 2022 micrometeoroid impact on segment C3 subsequently raised the high-order uncorrectable WFE term enough that the floor is now 59 nm rms.

[52] For example, in NIRCam's shortest filter F070W, the optical PSF pre-sampling is roughly 28 mas ∼ 0.9 pixels in the NIRCam short wavelength channel, but after sampling onto the detector pixel scale of 32 mas, the apparent PSF resolution becomes 40 mas ∼ 1.25 pixels in the NIRCam short wavelength channel.

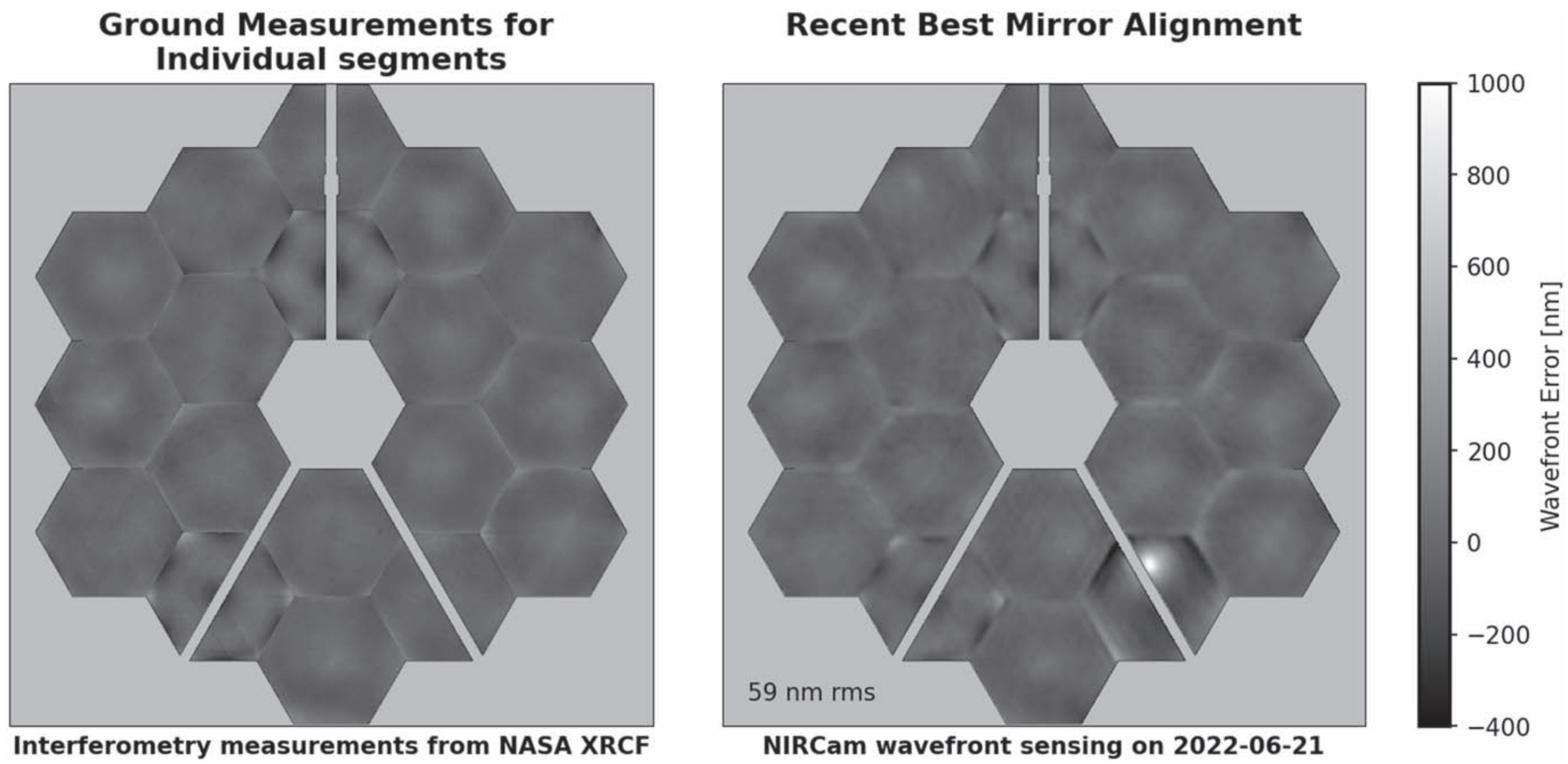

**Figure 3.** Wave front Sensing Measurements showing the quality of achieved mirror alignment on orbit. The telescope wave front error achieved in flight, shown in the right panel, closely tracks the as-polished surface figures of the individual segments, as measured during ground testing, shown in the left panel. JWST's wave front sensing and mirror control systems are working as intended, achieving optimal alignments within the ∼10 nm resolution of the sensing and control system and correcting as necessary to maintain that alignment. Data from PID 1163.

from room temperature to ∼45 K, the achieved wavefronts on each segment were just as expected.

### *4.2. Comparison to Optical Budget for the Telescope and Science Instruments*

JWST has top-level science requirements to achieve image quality with Strehl ratio greater than 0.8 in NIRCam at a wavelength of 2 $\mu$m (equivalent to wave front error of 150 nm rms) and MIRI at a wavelength of 5.6 $\mu$m (equivalent to 420 nm rms). These and other top-level requirements flowed into detailed optical performance budgets and lower-level requirements, such as a required telescope wave front error $\leqslant$131 nm rms over the fields of view of the science instruments, a required telescope stability of $\leqslant$54 nm rms over two weeks, and so on. Though line-of-sight image jitter is distinct from WFE, it is also tracked within the same budget via a computed equivalent WFE.

Figure 4 presents an abbreviated summary comparing measured performance in flight to those budgets at high level. Note that the measured values shown here are from commissioning at observatory "beginning of life," while the required and predicted values were derived for "end of life" after a notional 5 yr mission. At the observatory top level, the achieved WFE is well below predictions by ∼30%, and the remaining performance margin relative to requirements is large. All lower-level terms are at or below requirements values, with the sole exception of the high frequency WFE which was increased by the 2022 May micrometeoroid impact. Terms with particularly notable performance relative to requirements include the telescope stability (see Section 4.5), the line of sight pointing (Section 3.3), and the wave front quality of the science instruments (which are all significantly better than their requirements).

An initial assessment, which combined the measured beginning-of-life performance with model predictions for observatory aging in the L2 environment, predicts that JWST should meet its optical performance requirements for many years. The current largest source of uncertainty in models is the rate of mirror surface degradation from micrometeoroids, discussed below.

### *4.3. Shape of the Point-spread Function*

JWST's hexagonal aperture creates a characteristic diffraction pattern in its point spread functions, with six stronger diffraction spikes at 60° intervals created by the segment and aperture edges, plus two fainter horizontal spikes created by the vertical secondary mirror support. While these diffraction spikes can be visually dramatic in images which are deeply exposed or are plotted with log stretches, it is the case that the majority of light is focused into the PSF core (typically ∼66% within the first Airy ring).

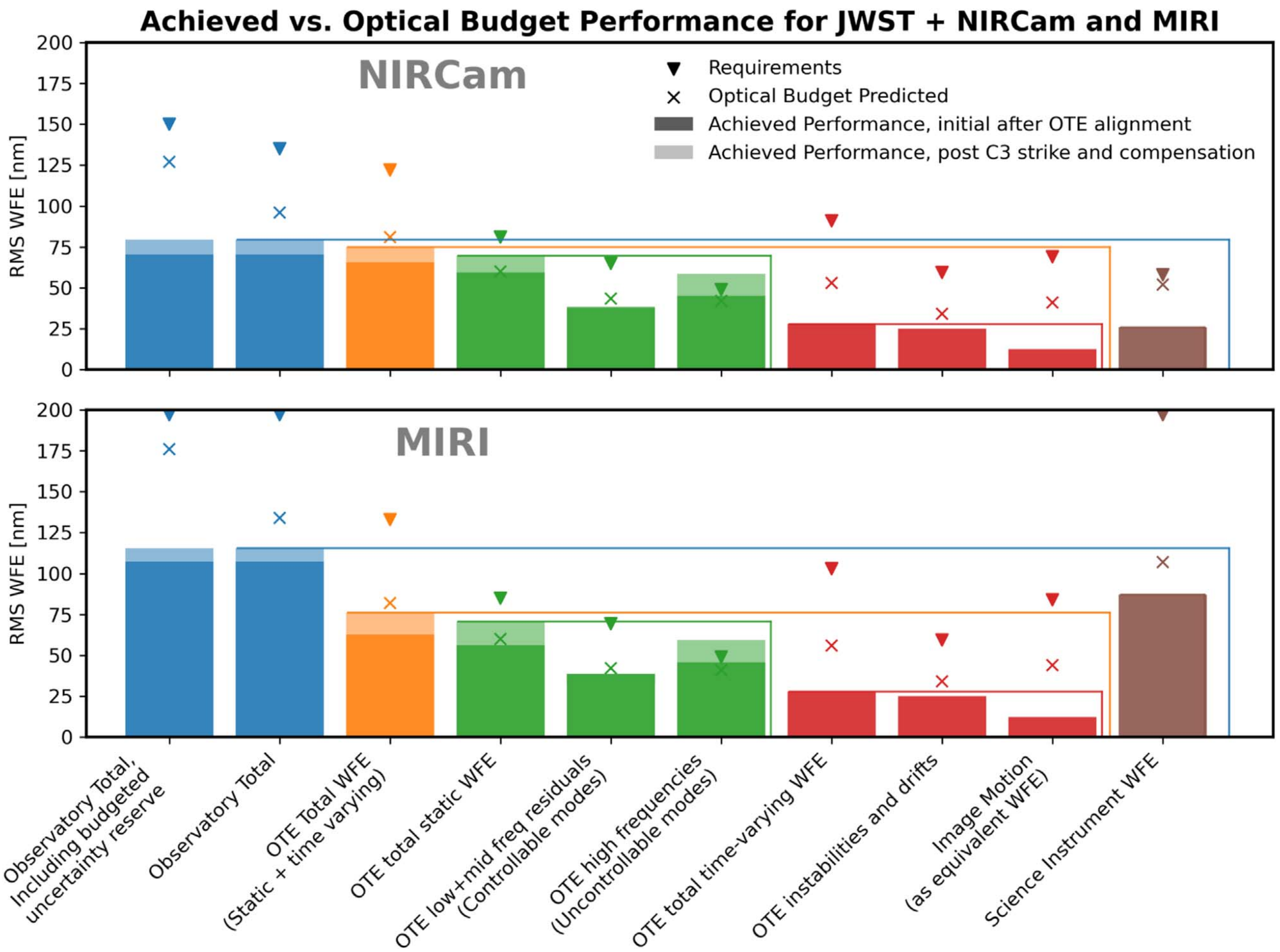

**Figure 4.** High-level summary representation of JWST optical performance for 2.0 and 5.6 $\mu$m. For each contribution to wave front error, triangles represent the required level, Xs mark the pre-launch optical budget predicted levels, and bars indicate the measured performance. All Optical Telescope Element (OTE) values shown are from on-orbit measurements. Science instrument WFE values shown are for typical field points (median SI image quality) in NIRCam and MIRI, as measured on the ground during ISIM CV3 testing. The colored lines depict which terms sum together; for instance the OTE total WFE is the RSS sum of the OTE static WFE and OTE time-varying WFE. Shaded portions of the bars indicate the delta in performance from the 2022 May micrometeoroid strike on segment C3.

Compared to the diffraction patterns from circular apertures which many astronomers are more familiar with, such as Hubble, the hexagonal geometry of JWST concentrates wide-angle diffracted light more strongly into the diffraction spikes, while the areas between those spikes are relatively darker. For bright sources the diffraction spikes can be seen to a separation of many arc minutes, falling off as $\sim R^{-1.5}$, including diffraction spikes from sources outside of an instrument's field of view.

Note that the position angles of the six bright diffraction features are different between the PSF core region (2–5 $\lambda/D$, dominated by diffraction from the overall outer hexagonal outline) and the outer wings ($>5$ $\lambda/D$, dominated by diffraction from the individual segments). See Figure 2. This can be understood intuitively from the pupil geometry: when assembling a larger hexagon from smaller hexagons, the outline of the larger hexagon which is created is rotated 30 deg relative to the smaller ones.

*Additional PSF details per instrument*:

1. MIRI imager and MRS PSFs, particularly at wavelengths $\leqslant 10$ $\mu$m, show additional diffraction spikes in vertical and horizontal directions in the instrument coordinate frame, called the "cruciform" or "cross artifact." These arise due to diffraction that occurs internal to the detector at the detector electrical contacts. This effect was seen in Spitzer's similar Si:As detectors, and was modeled and expected for MIRI (Gaspar et al. 2021). Improved models based on flight data are able to reproduce this artifact in detail including the field dependence. Pipeline steps are being developed to compensate for its impact on MIRI MRS data cubes.

2. NIRISS imaging PSFs at wavelengths $\geqslant 2.7\,\mu$m have stronger diffraction spikes due to additional pupil obscuration from the CLEARP pupil wheel position which must be used with those filters.
3. NIRISS imager PSFs at wavelengths $\leqslant 2.0\,\mu$m show additional anomalous or "extra" diffraction spikes, which are more diffuse than the ordinary spikes and have a position angle that varies strongly with field position. The intensity of the anomalous spikes decreases with wavelength, in F090W containing ~70% of the flux of the normal vertical diffraction spike, becoming barely detectable by F200W. These spikes are now understood as due to diamond-turning tool marks on off-axis mirrors within NIRISS. Retroactive examination of some ground test data also shows their presence at lower S/N. A similar effect is also seen in FGS PSFs.
4. The rotation of NIRSpec in the focal plane means that, in the detector coordinate system, NIRSpec PSFs are rotated by 139 deg relative to the other instruments. MIRI is similarly rotated 5 deg.
5. Instrument modes with specialized optical components, such as NIRCam coronagraphy, MIRI coronagraphy, NIRISS aperture masking interferometry, and NIRISS SOSS, each have their own PSF properties, as described in Girard et al. (2022), Kammerer et al. (2022), Kammerer et al. (2023), Boccaletti et al. (2022). These observed PSFs are in general highly consistent with model predictions.

### *4.4. Transmission and Contamination*

The telescope's effective area, the product of the telescope area and its transmission, is a key optical performance parameter that was tracked through development and characterized on orbit. The JWST unobscured telescope area was required to be $>25\ \mathrm{m}^2$. The measured value using the NIRCam pupil imaging lens was $25.44\ \mathrm{m}^2$. The telescope's wavelength-dependent transmission ranged from 0.786 at $0.8\,\mu$m to 0.933 at $28\,\mu$m, better than requirements at each wavelength. Even though the JWST optics spent significantly more time in ground facilities than originally anticipated, these high transmissions were maintained with careful control of contamination throughout the integration and test phases and during preparation for launch. Brush cleaning was also carried out on the telescope primary mirror segments and secondary mirror to remove particulates.

### *4.5. Optical Stability on Different Timescales*

Commissioning observations characterized the telescope's optical stability and variations on multiple timescales. Highly precise wave front sensing allows measurement of very small changes (<10 nm). The overall amplitudes of variation are close to predicted values, however some timescales are faster than expected. These variations have small but measurable effects on PSF properties, which should be taken into account for measurements at the highest precision, but will be negligible for many science cases (similar to the effect of the "breathing" variations seen in Hubble PSFs on orbital timescales).

Overall stability: During the last month of commissioning, 2022 June, the change in wave front measured between successive observations roughly 2 days apart was typically $\leqslant$25 nm rms, and frequently less than half that (range 8–50 nm rms per 2 days). During this time period, the observatory was conducting a wide range of commissioning and Early Release Observations typical of science activities, so this level of stability should be representative of what can be expected during Cycle 1. Indeed, during the first half of Cycle 1 the median change in wavefront between successive observations was only 10 nm rms, just slightly above the typical wavefront sensing measurement uncertainty of 7 nm rms.

Though we provide details below on observed variations over time, we wish to emphasize that these are *small* variations. The absolute amplitude of drifts seen between successive wave front monitoring visits is comparable to the Hubble Space Telescope's ~18 nm rms of focus variation typical on orbital timescales (see e.g., Lallo 2012). Yet for JWST the variations are observed to mostly occur on significantly longer timescales, and the effect on images at longer wavelengths is correspondingly reduced. (Hubble is typically stable to $\lambda/30$ at $\lambda = 0.5\,\mu$m over 90 minutes; JWST is often stable to $\lambda/100$ at $\lambda = 2\,\mu$m over 2 days.) Also similar to Hubble, the amplitude of wave front variations over time is comparable to and generally less than the field-dependent variations within a given instrument.

Factors contributing to stability levels:The telescope is very thermally stable, but small changes in equilibrium temperature (<0.1 K, within the temperature sensor noise) can still occur in response to changes in attitude with respect to the Sun. Variations from instrument heat sources can also affect telescope structures. A thermal stability exercise measured these effects by moving between the extremes of JWST's field of view, with a week-long soak at "cold" attitude ($-40°$ pitch relative to the Sun) sandwiched between an equal amount of time at the "hot" attitude (0° pitch); this test is intentionally more stressing than typical attitude profiles during science operations. The thermal slew exercise (PIDs 1445, 1446) included a wide range of wave front sensing, imaging, and sensor telemetry investigations to characterize many aspects of JWST's performance on multiple timescales. Modes of variation observed with JWST include the following.

#### 4.5.1. Telescope Backplane Thermal Distortion

After backing out the contributions from tilt events (see below), the observed drift in wave front during the thermal slew test could be well fit as a double exponential. The largest and slowest term arises from thermal deformation of the observatory backplane in response to the small change in equilibrium temperature. In the thermal slew test this was measured to be ∼18 nm with a time constant of 1.5–2 days. In comparison the preflight prediction from integrated modeling was 14.6 nm with a time constant of 5–6 days predicted at observatory beginning of life, versus requirements of 54 nm rms. The dominant mode in this drift is zero-degree astigmatism, essentially a bending mode of the primary outer segment "wings" relative to the center section, in accordance with predictions.

#### 4.5.2. Telescope Soft Structure Thermal Distortion

The second term in the double exponential fit represents wave front drift which arises from soft structures. The black Kapton stray-light-blocking "frill" around the outside of the primary can place mechanical stress on its points of attachment when it thermally expands and contracts. The lower thermal mass of the frill causes this to have a shorter characteristic timescale. The observed effect in the thermal slew test was an exponential with 4.45 nm amplitude and time constant of 0.77 hr.[53] mac-fn7 In comparison, the preflight prediction was 8.6 nm with a time constant of 8–10 hr.

#### 4.5.3. Fast Oscillations from Heaters

On short timescales (2–4 minutes), the thermal cycling of heaters in the ISIM Electronics Compartment (IEC) induces small forces on the telescope backplane structure. The observed result is a semi-periodic variation primarily in astigmatism with amplitude of about 2.5 nm, which is closely consistent with preflight predictions. The exact amplitude and timescale vary as the cycles of several different heaters beat together. This oscillation is sufficiently small and rapid that it has thus far only been measured in special high-cadence differential wave front sensing measurements which achieve sub-nanometer precision. This effect is expected to be negligible for the vast majority of science observations.

#### 4.5.4. Segment Tilt Events and other Drifts

"Tilt events" are occasional abrupt changes in position of an individual segment which are seen to occur from time to time, with varying amplitudes of a few nanometers to tens of nanometers. See Figure 23 of McElwain et al. (2023) (this issue). These are hypothesized to be due to structural microdynamics within the telescope (e.g., localized relaxations of stiction or microstresses within the backplane structure). This hypothesis is supported by a few cases in which measurable stress relief was detected by telescope wing latch strain sensors at the time of tilt events. Fast wave front sensing measurements constrain the timescale for some tilt events to be <10 s; these are effectively instantaneous step functions in mirror position.

Since a few such events were first seen during the telescope cryovacuum testing on the ground, it was expected these might be seen in flight, particularly initially after cooldown. Tilt events proved to be not infrequent during both commissioning and early Cycle 1, though the rate has decreased over time. The rate of tilt events continued to decrease through the first half of Cycle 1, and as of this writing in 2023 February there have been no additional larger tilt events since 2022 October. During 2022 April–May there were periods in which small tilt events occurred as frequently as two to three per day, but by later 2022 June there were weeks with few or no tilt events. Figure 23 of McElwain et al. (2023) (this issue) shows that tilt events became increasingly rare during the first five months of science operations (2022 July through November). The expectation that tilt events would be small and would stabilize as the microstresses are relieved seems to be in line with the reduction we are seeing and the fact that they are small.

Tilt events may cause small but observable changes in PSFs during some science observations. The resulting change to encircled energy is very small (small fractions of a percent) and much smaller than encircled energy stability requirements. However it can be significant for observations at very high precision. For instance tilt events were observed in the NIRCam and NIRISS time series commissioning tests, and caused abrupt small step-function offsets in the initial measured flux analysis (for example, see Section 6.1.2). Flux jumps caused by occasional tilt events can be calibrated as a function of aperture size or wavelength and/or be included as a parameter in the transit fit model. The time sequence of FGS image data and centroids obtained at 15.6 Hz during all science observations may also be a useful diagnostic channel for some tilt events.

Not all changes in segment position occur on the rapid timescale of tilt events; at times, slow drifts of segment positions over hours or days have been observed as well. Understanding of the structural dynamics of this segmented telescope in space will continue to improve with time.

### 4.6. Routine Wave front Sensing and Control

Since the completion of telescope alignment, regular wave front sensing and control has monitored and maintained telescope alignment, as will continue throughout the mission. During science operations, wave front sensing measurements are conducted every two days (roughly, with flexible cadence around science observations). The resulting measurements of mirror alignment state versus time are automatically made available in the MAST archive; software for making use of these measurements to produce time-dependent PSF models is now included in the WebbPSF package.[54]

[53] While one might assume the amplitude of this effect to be independent of slew direction, this was not observed to be the case: after the reverse slew from cold to hot attitude, a similar exponential drift was not seen, and instead the wave front was stable to within 3 nm over the 6 hr measurement period. Due to test duration it was not possible to repeat these measurements for further investigation of under what circumstances the frill term does or does not manifest.

Wavefront control is nominally applied whenever the total wavefront error (telescope + NIRCam, short wave channel, at the reference field point) reaches 80 nm rms due to drift or any other instability. Over the period 2022 November through 2023 January, this total has varied over a narrow range between about 65–77 nm rms, with only two wavefront corrections needed.

Because of the every-two-days cadence of wave front sensing measurements, wave front states at times between those measurements are not directly measured (e.g., if a tilt event or micrometeoroid impact occurs within some 2 days, its exact time of occurrence is not measured). Some science modes do allow for more frequent wave front measurements, in particular NIRCam time series using the weak lens in the short wave channel, and to some extent the NIRISS SOSS mode.

### *4.7. Micrometeoroids*

Inevitably, any spacecraft will encounter micrometeoroids. The part of JWST that is most vulnerable is the primary mirror. Some of the resulting wavefront degradation from micrometeoroid strikes is correctable through regular wavefront control, while some of it comprises high frequency terms that cannot be corrected; this latter, cumulative damage was incorporated into the prelaunch JWST wavefront error budget (Feinberg et al. 2022). Over the period 2022 March 11 to 2023 January 12, wavefront sensing recorded a total of 25 localized surface deformations on the primary mirror that are attributed to impact by micrometeoroids (a hit rate of 2.5 per month.) With one major exception, after correction these micrometeoroid strikes have had no measurable impact on the overall wavefront error.

The outlier is the micrometeoroid hit to segment C3 in the period 2022 May 22–24, which caused significant uncorrectable change in the overall figure of that segment. However, the effect was small at the full telescope level because only a small portion of the telescope area was affected. After two subsequent realignment steps, the telescope was aligned to a minimum of 59 nm rms, which is about 9 nm rms above the previous best wave front error rms values.[55] Since the NIRCam shortwave channel has the least wavefront error of all the modes (see Table 2 of McElwain et al. (2023), this issue), the increase in wavefront error for all other instruments and modes from the C3 strike should be comparable to or less than NIRCam.

Images of the telescope optics using the NIRCam pupil imaging lens can reveal smaller impacts below the threshold detectable by wave front sensing. Comparison of pupil images taken 23 February and 2022 May 26 show evidence for 19 such minor strikes over that 92 days period. Regular monitoring of the pupil may help constrain the micrometeoroid hit rate and power spectrum. Micrometeoroid impacts should also slightly lower the telescope throughput; this effect is not yet measurable.

It appears the 2022 May hit to segment C3 was a fairly rare event. Still an open question is how rare — every year versus every few years. Ten such large hits would degrade the mirror such that it is no longer diffraction limited at lambda < 2 micron. Therefore, after further investigation of the micrometeoroid population (see Section 6.2 of McElwain et al. (2023), this issue), the JWST Project has decided, starting in Cycle 2, to limit the amount of time the telescope spends pointed toward the direction of orbital motion, as those directions statistically have higher micrometeoroid rates and energies.

## 5. Backgrounds, Stray Light, and Scattered Light

### *5.1. Backgrounds*

The level of background emission is critical to the depth of many JWST imaging and low-spectral-resolution spectroscopic observations. In addition to the unavoidable in-field backgrounds from our solar system and our galaxy, as an unbaffled telescope with a non-zero temperature, JWST sees two additional sources of background: (a) the astronomical stray light, mainly affecting the near-infrared wavelengths, in which light is scattered into the field of view; and (b) thermal self-emission from the glowing mirrors, sunshield, and other observatory components, which affects the mid-infrared. Before launch, the predicted levels of these components carried uncertainties of ∼20% and ∼50%, respectively. These backgrounds, now measured from commissioning and early science data, are described in a companion PASP paper (Rigby et al. 2023). Here, we briefly summarize the key results.

First, the astronomical stray light is observed at levels considerably below the requirements, and 20% lower than the pre-launch predictions. As a result, deep fields at high ecliptic latitude will go deeper faster than expected for wavelengths <5 $\mu$m.

In the mid-infrared, the main JWST background is a combination of thermal emission from the primary mirror and scattered thermal emission from the sunshield and other parts of the spacecraft. The measured background spectrum is very close to the predicted thermal spectrum that was incorporated into the exposure time calculator before launch, with somewhat higher backgrounds at at 5–15 $\mu$m. The measured thermal background at 10 and 20 $\mu$m is close to the requirements values. The mid-infrared backgrounds in the F2550W filter are variable at the level of 7% over timescales of weeks to months, which is less variability than expected, due to greater than expected thermal stability of the primary mirrors.

[54] Available at https://www.stsci.edu/jwst/science-planning/proposal-planning-toolbox/psf-simulation-tool.

[55] The impact raised the wave front error of segment C3 from 56 to 280 nm rms. Mirror commanding to adjust segment position and curvature reduced this error to 178 nm rms. This, after dividing by area and adding in quadrature to the other sources of WFE in the telescope, results in ∼9 nm rms increase to the total telescope wave front error.

Given these measurements, JWST is indeed, as it was designed to be, limited by the irreducible astronomical background emission, not by stray light or its own self-emission, for all wavelengths <12.5 $\mu$m. The control of stray light and thermal emission is an engineering triumph that will translate into substantially better than expected scientific performance for many applications.

### 5.2. Scattered Light Features

During commissioning, a few unexpected stray light features were discovered, characterized and understood, and mitigation plans were developed. Early science operations have shown the success of these mitigations.

There exists a stray light path, termed the "rogue path," that bypasses the primary and secondary mirrors, directly passes through the aft optics system (AOS) aperture just over the back of the fine steering mirror, and reaches the science instrument pick off mirrors (Lightsey et al. 2014). While the rogue path was anticipated, and steps taken to block it to the extent possible, the NIRCam and NIRSS instruments at times show some unexpected scattered light features which have been identified as being due to this path. The resulting features are relatively faint, but for some observations such as deep fields these could be significant noise terms if not mitigated. More information may be found at the JDox Data Features and Image Artifacts page. Here we summarize these features and their mitigations.

#### 5.2.1 Wisps

A few percent of the pixels in four of the eight NIRCam short wavelength detectors show small, faint, diffuse features that are termed "wisps." See Figure 5. In the B4 detector wisps are always present, with variable brightness that is typically about 10% of the zodiacal background. Fainter wisps have also been seen in detectors A3, A4, and B3. Wisps occur at fixed detector positions. The origin of wisps has been traced to reflections from the upper strut that supports the secondary mirror. Wisps have not been seen in the NIRCam long wavelength channel.

#### 5.2.2 Claws

A minority of NIRCam short wavelength channel images from commissioning showed "claws"—a faint diffuse pattern of scattered light features that moves across the detector coordinates as the telescope dithers. See Figure 5. Claws are rare, but when present, have a brightness of about 10% of the zodiacal background. When present, claws occur primarily in the NRCA1 or NRCB4 detectors, affecting roughly 5% of the pixels on those detectors. Since claws move in detector coordinates, they are more difficult to subtract off than wisps. Claws arise when a bright star is located in the rogue path susceptibility region.

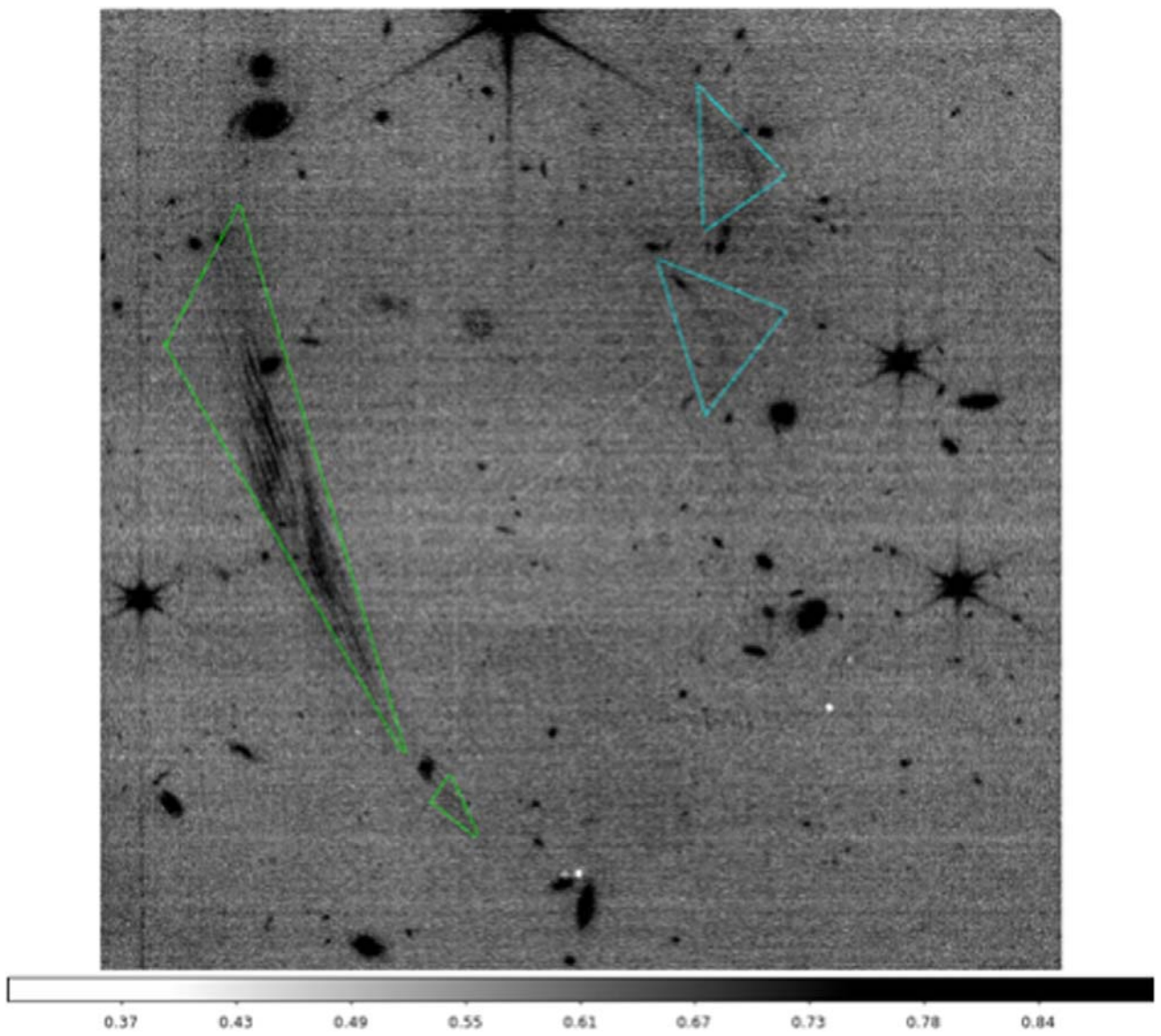

**Figure 5.** The NIRCam scattered light features "the claws" and "the wisps." The claws are marked by green triangles, and the wisps by cyan triangles. Roughly 4% of the pixels in the detector B4 are involved in a claw, and 1.5% in a wisp. Data are from detector B4 of the NIRCam short wavelength channel. Of the 8 short-wavelength detectors in NIRCam, detector B4 is the one most affected by wisps, and one of the two most affected by claws. Data are from program PID 01063, in which bright star X Cancri ($K$ = 0.25 mag) was inside the claws susceptibility zone. Image used is jw01063142001_02101_00001_nrcb4_rateints.fits. The detector covers 63″ on a side.

#### 5.2.3 The Lightsaber

Some NIRISS images show a narrow band of excess stray light running almost horizontally across the detector. Dubbed "the lightsaber," this light originates from the rogue path (see NIRCam claws above), grazes off the NIRISS entrance housing wall and then experiences double reflections off two mirrors inside NIRISS. See Figure 6. The light comes from a susceptibility region mapped and modeled to be far away from the NIRISS field-of-view ($+2^\circ\!.0 < V2 < +5^\circ\!.0$, $+12^\circ\!.4 < V3 < +12^\circ\!.8$). The zodiacal light and any stars in this region contribute to the intensity. When the light is dominated by the zodiacal light the observed brightness is typically up to 1.5% per pixel of the in-field background. When bright ($H_{\rm Vega} \sim 0$) stars are in that region, the observed brightness is up to 10% per pixel. The lightsaber from bright stars is somewhat

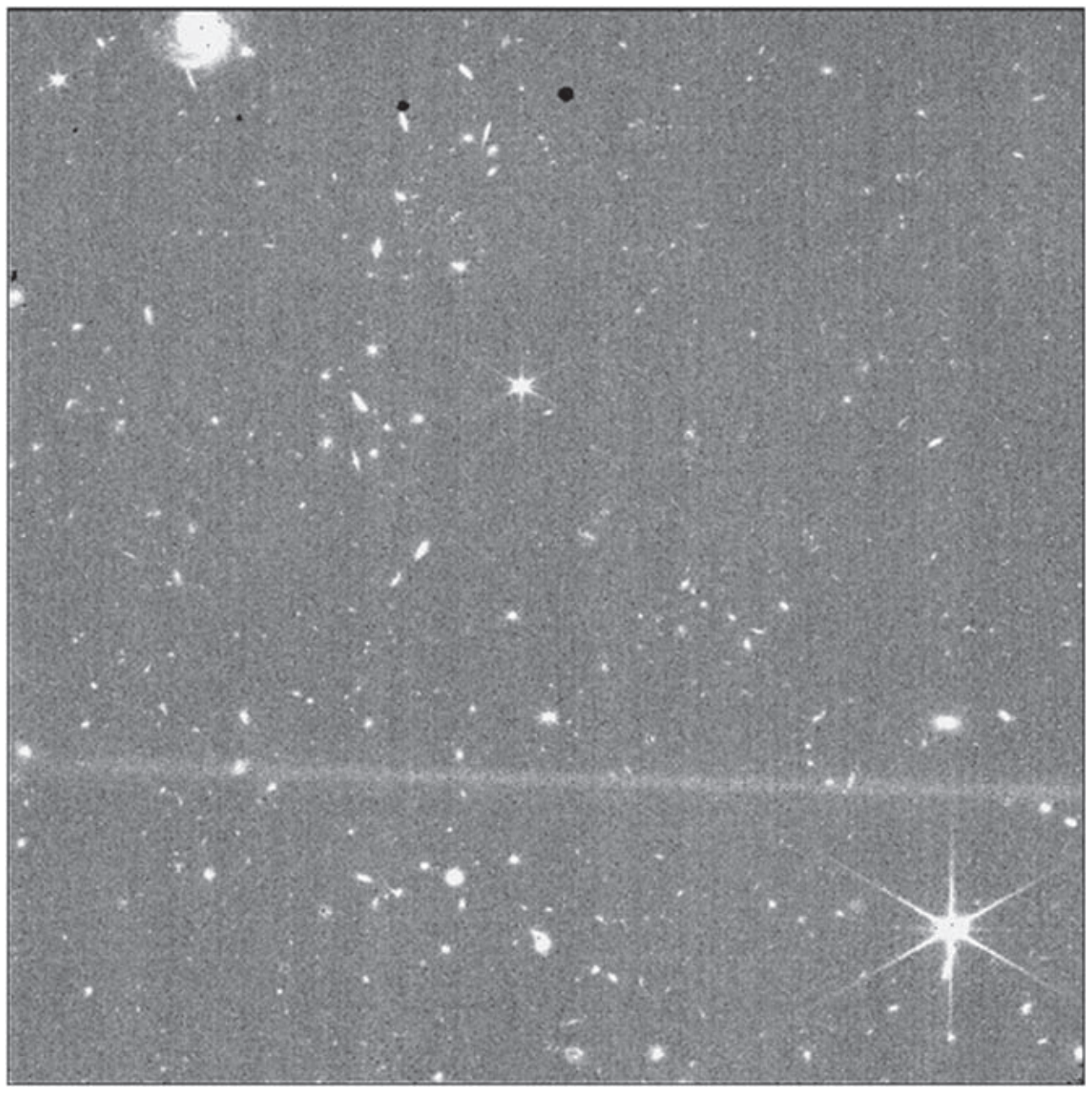

**Figure 6.** The NIRISS lightsaber. This image in the F150W filter shows a prominent lightsaber stray light feature running almost horizontally across the lower part of the image. For this pointing there is a bright star in the rogue path susceptibility region. The field of view of the detector is 2$'$.2 on a side. Data from PID 1063.

narrower and shifts position within the broader band caused by the zodiacal light.

#### *5.2.4 Mitigation of the Stray Light Features*

Stray light modeling associated the claws and lightsaber with the presence of bright (0–1st Vega mag) stars in the "rogue path" susceptibility zone: a region of the field of view located several degrees off the telescope boresight (10° for NIRCam, 13° for NIRISS). Light from bright stars in this zone can bypass the telescope optical train, enter the SIs through the SI pickoff mirrors, then bounce from non-optical surfaces within the science instruments to the detectors. During commissioning, we directly mapped the extent of the rogue path susceptibility zones for both NIRCam and NIRISS by moving a bright star through the susceptibility zone.

The claws and lightsaber may be avoided simply by ensuring observations do not place very bright stars in the susceptibility zone. In early science observations, manual checks of scheduled observations by STScI staff have largely prevented recurrence. STScI plans to update the Astronomers' Proposal Tool to alert users when planned observations would place bright stars in the susceptibility zone, to give users advance predictions of the artifacts and allow replanning observations to avoid them. The much fainter but ever-present lightsaber due to zodiacal light has been modeled, and can be scaled and subtracted from images or low-resolution grism spectra. More information on this and other NIRISS features may be found at the Data Features and Image Artifacts JDox page.

The "wisp" features in NIRCam arise from a different optical path: stray light that reflects from the upper strut which supports the secondary mirror. These are fixed in position; the NIRCam team has demonstrated that they subtract out well (see Rieke et al. 2023, this issue).

## *6. Science Instrument Performance*

JWST has 17 science instrument modes. Near the end of commissioning, the science performance of each mode in turn was reviewed against criteria developed pre-launch for such parameters as sensitivity, image quality, wavelength calibration, astrometric calibration, ghosts, stability, and so on. As of 2022 July 10, all 17 of the 17 modes have been approved to begin science operations. With the performance of the instruments being uniformly excellent versus requirements and typically also better than pre-flight estimates, the assessments of modes as being ready for Cycle 1 science have been quite straightforward. Below we summarize the performance and any known issues for each instrument mode.

A key result of science instrument commissioning is that overall, the JWST science instruments have substantially better sensitivity than was predicted pre-launch. This result is due to higher science instrument throughput, sharper point spread functions, cleaner mirrors, and lower levels of near-infrared stray light background compared to pre-launch expectations. The exposure time calculator (ETC) and its underlying Pandeia engine were overhauled in v2.0 (released 2022 December) to reflect on-orbit performance, timed to support the Cycle 2 Call for Proposals. The JDox documentation has been similarly updated. The ETC and its Pandeia engine are the definitive reference as to sensitivity. For now, we quote several representative measurements and calculations that were determined during commissioning.

In Figures 7 and 8, we summarize the sensitivity of JWST in two common modes: imaging and emission line spectroscopy, and compare to previous and current observatories. We convert from the continuum sensitivity calculated by Pandeia to the more intuitive limiting emission line flux, using the following equation:

$$f_{\rm line} = f_{\rm cont}^{\rm pix}\, 10^{-23} w \sqrt{l}\,,$$

where $f_{\rm line}$ is the limiting emission line flux in erg s$^{-1}$ cm$^{-2}$, $f_{\rm cont}^{\rm pix}$ is the per-pixel continuum sensitivity in Janskies, w is the pixel width in Hz, and $l$ is the line width in pixels.

Across all instruments, JWST has multiple time series observation (TSO) modes to support observations of transiting exoplanets. During commissioning, observations of the exoplanet HAT-P-14-b were obtained with three NIR spectroscopic modes—the NIRCam grism time series mode, the NIRISS SOSS mode, and the NIRSpec bright object time series mode—to enable cross-comparison between instruments and assessment of astrophysical versus instrumental systematics. This target was chosen because, as a massive planet with a high surface gravity, it is expected to have a flat transmission spectrum. It also has a relatively bright host star ($K_{s,\mathrm{Vega}} = 8.9$). In addition the transiting exoplanet L 168-9 b was observed with the MIRI LRS time series mode. Results are given in the subsections below for each instrument; the overall picture is that JWST is returning precise transit spectra with minimal processing.

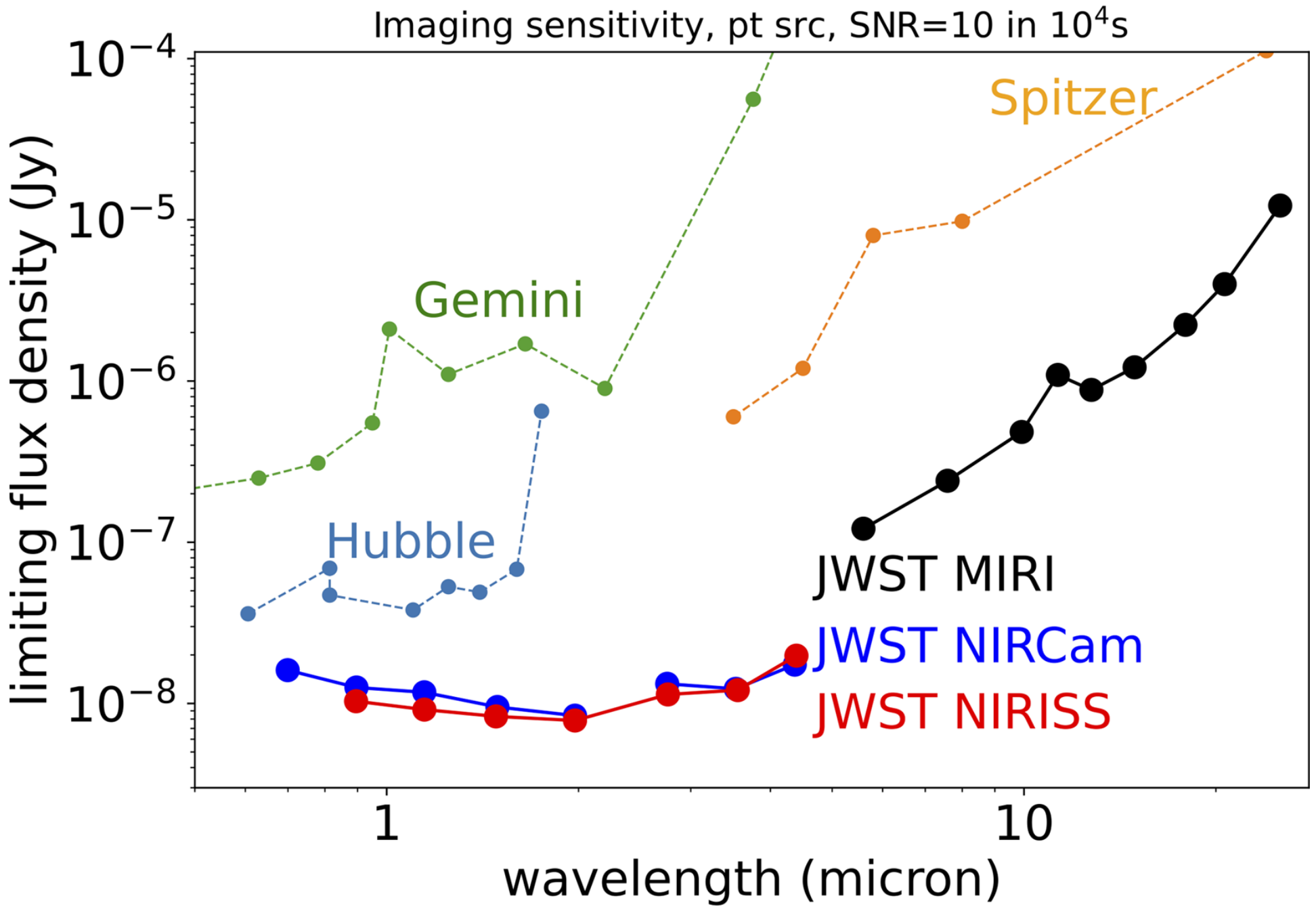

**Figure 7.** Imaging sensitivity for JWST. The *Y*-axis shows limiting flux density: the faintest point source that can be detected at $S/N = 10$ in an integration time of ten thousand seconds, in units of Janskies. The *X*-axis is wavelength in units of microns. JWST instruments are shown by the points connected by solid lines, color-coded by instrument as NIRISS (red), NIRCam (blue), MIRI (black). JWST brings two orders of magnitude improvement in imaging sensitivity at 2–3 $\mu$m. Actual JWST sensitivity was calculated using Pandeia v2.0, which is the version of the exposure time calculator engine that was released to support the Cycle 2 call for proposals; calculated sensitivities in this released version reflect on-orbit performance as characterized during commissioning. For comparison, comparable sensitivities are shown for other observatories, with points connected by dashed lines: Hubble (WFC3, ACS, and NICMOS instruments); Gemini (GMOS and NIRI instruments); and Spitzer (IRAC and MIPS instruments). Plotted sensitivities for JWST and the comparison observatories are included as supplementary data tables, which also describe computation methods for the sensitivity of these comparison instruments are given in the supplementary information.

Mode readiness reviews also included the documentation of any remaining work needed to enable particular use cases (e.g., nudging a subarray location slightly to not clip a spectrum; updating an astrometric file needed for target acquisition) or the identification of performance features that observers should be made aware of before final planning of their observations (e.g., alerts regarding faster-than-expected saturation due to higher-than-expected throughput, or warnings regarding stray light features such as the "claws" and "lightsaber," with tips for mitigation). These issues were captured as "liens" on the mode readiness that must be addressed for some specific observing programs (or that are being addressed by the operations team in the near term). Liens have not prevented any Cycle 1 observations from executing, nor have any observing windows been missed due to a lien. Liens have caused delays in scheduling some programs, and some have required work-arounds to make some programs executable before a permanent fix for the lien was ready.

Observations not affected by liens are ready for immediate insertion into the observing plan. Following the first several mode readiness reviews, cycle 1 observations of early release science targets began on 2022 June 20. General Observer and Guaranteed Time Observer cycle 1 programs followed.

### *6.1. NIRCam Performance*

The Near Infrared Camera (NIRCam) instrument is described in the companion PASP special issue paper Rieke et al. (2023). Here, we summarize the science performance of NIRCam as characterized during commissioning.

#### *6.1.1. NIRCam Imaging*

The throughput of NIRCam meets or slightly exceeds pre-launch expectations for all but a few of the filters in the short wavelength channel. In the long wavelength channel, the throughput is systematically 20% higher than expected for most filters. Figure 9 shows the throughput of NIRCam compared to what was assumed in the pre-launch version of the ETC.

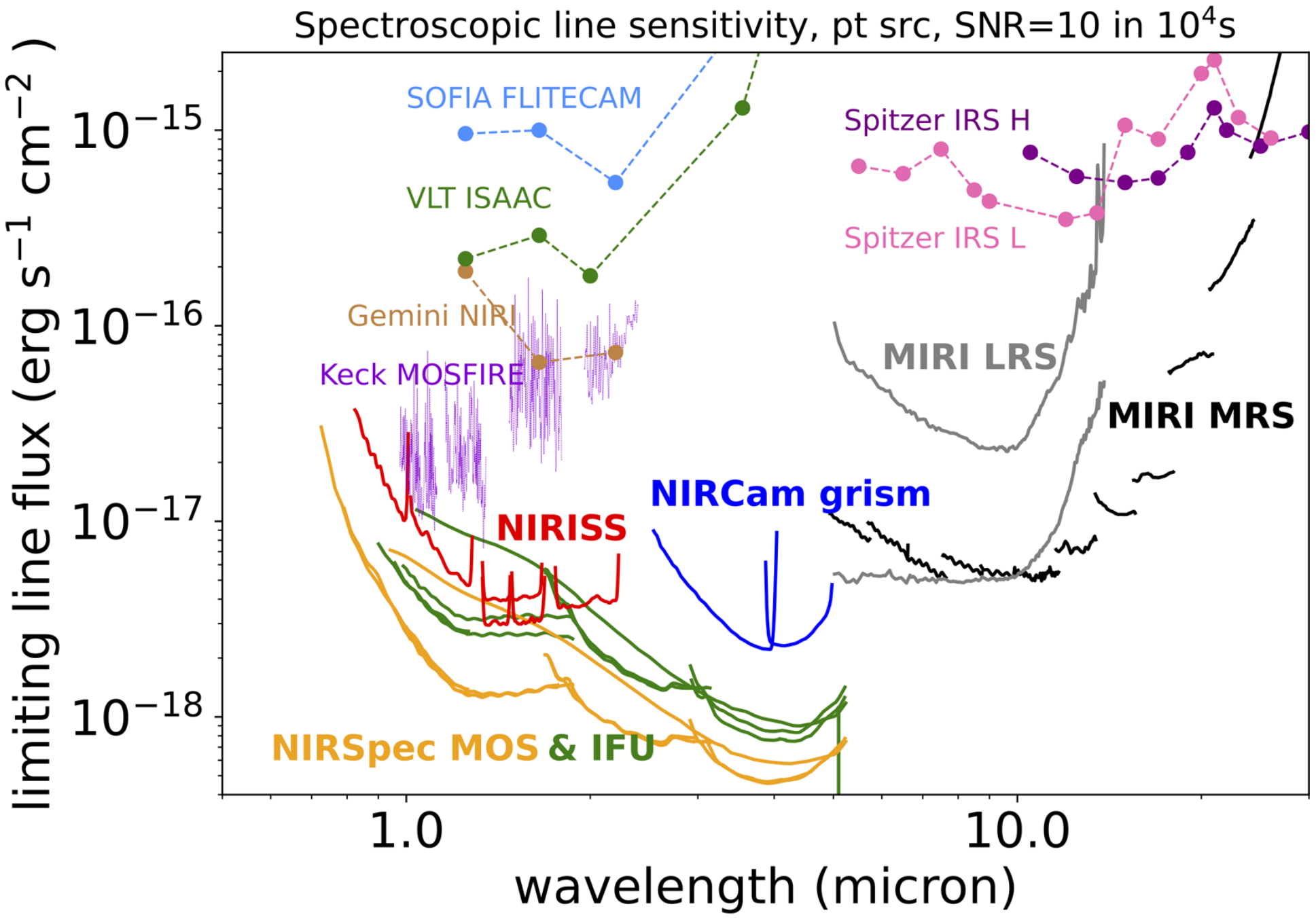

**Figure 8.** Spectroscopic sensitivity for JWST. The $Y$-axis shows the limiting line flux, which is the flux of the faintest narrow (spectrally unresolved) emission line in a point source that can be detected at $S/N = 10$ in an integration time of ten thousands seconds. The JWST instrument sensitivities are plotted in bold lines and labeled in boldface. JWST brings dramatic improvement in spectroscopic sensitivity and spectral resolution compared to previous observatories. As in Figure 7, calculations for the JWST instruments were done using Pandeia v2.0. For comparison, sensitivities are plotted (thin dashed lines) for other observatories: SOFIA (FLITECAM; sensitivity scaled from their exposure time calculator), Gemini (NIRI instrument; sensitivity from instrument website), VLT (ISAAC instrument, sensitivity from their ETC), Keck (MOSFIRE instrument; sensitivity from Wirth et al. 2015), Spitzer (the IRS instrument, for the "$L$" or low-resolution ($R$ = 60–120) gratings and the "$H$" high ($R$ = 600) resolution gratings; sensitivity from the SPEC-PET calculator). Plotted sensitivities for JWST and the comparison observatories are included as supplementary data tables, which include more information on how the comparison sensitivites were calculated.

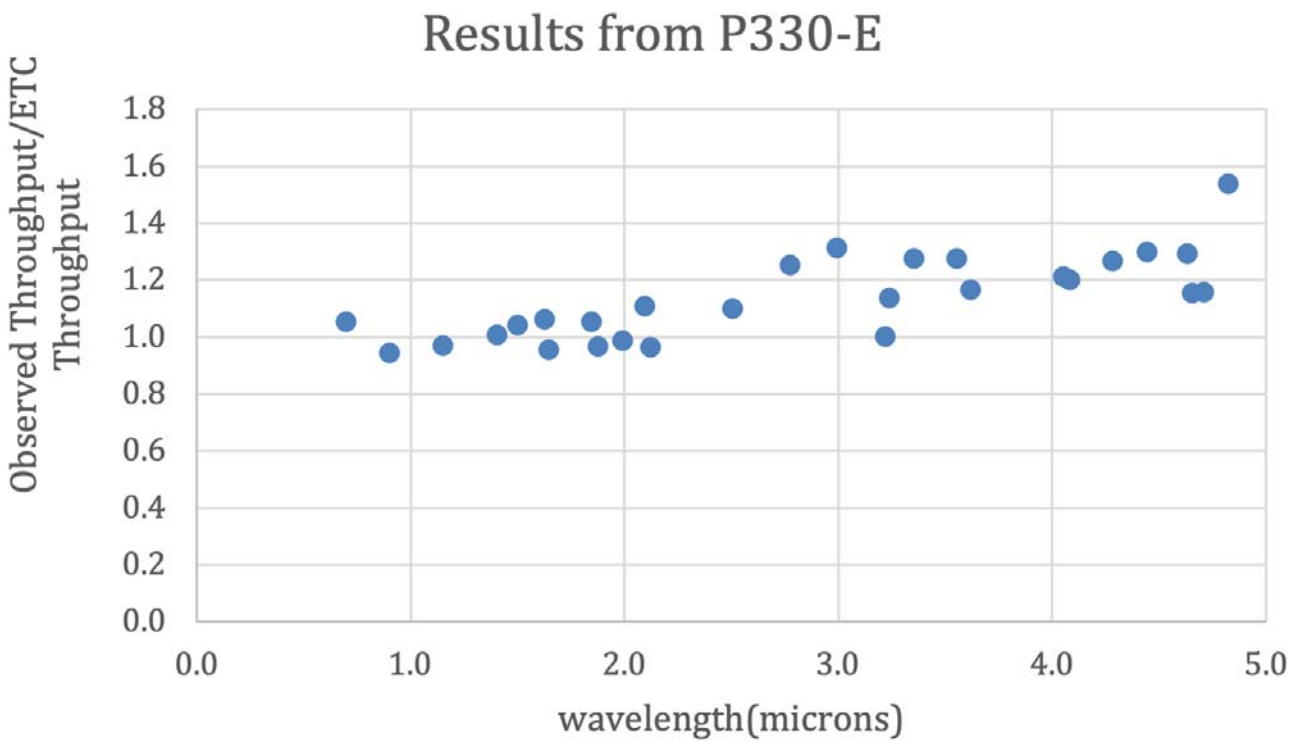

**Figure 9.** NIRCam imaging throughput compared to what was assumed in the pre-launch ETC. For most filters, the observed throughput is higher than the pre-launch expectations. Data are from spectrophotometric standard star P330-E observed in program PID 1074.

**Table 3**
NIRCam Limiting Point Source Sensitivity

| Wavelength ($\mu$m) | 2 | 3.5 |
|---|---|---|
| filter | F200W | F356W |
| Requirement (nJy) | 11.4 | 13.8 |
| ETC prediction (nJy) | 10 | 14.1 |
| Actual (nJy) | 6.2 | 8.9 |

**Note.** What is quoted is the faintest flux density (in nanojanskies) that can be detected at $S/N = 10$ in 10,000 s, for imaging in broad-band filters, assuming a background that is 1.2 times the minimum zodiacal light level. Smaller numbers are better. The equation to convert from flux density $f$ in nJy to $AB$ magnitudes is: $mAB = -2.5\ \log_{10}[f(\mathrm{nJy}) \times 10^{-32}]\ -48.57$. Actual sensitivities are from Table 2 of Rieke et al. (2023), this issue).

The point-spread function is better than expected, as parameterized by encircled energy or full width at half maximum. The photometric stability is stable to at least 4%, and is likely much better. The residual astrometric errors are 2–4 mas per filter across a detector. Optical ghosts are consistent with expectations from ground tests. Some scattered light features are seen (the "claws" and "wisps," described in Section 5.2), which may have some impact on deep imaging.

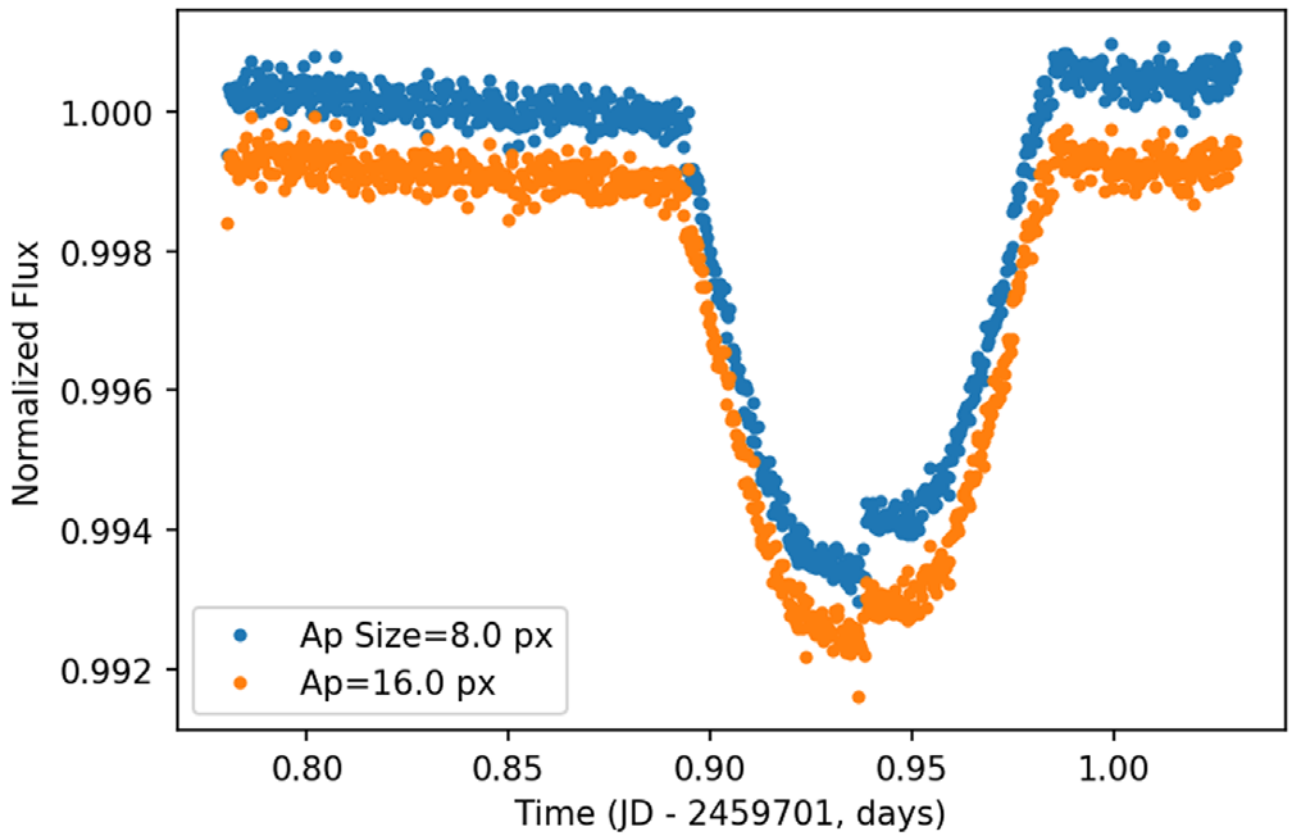

**Figure 10.** The summed broadband (2.4–4.0 $\mu$m) grism light curve from the HAT-P-14 observation. The lightcurve from the NIRCam long wavelength channel shows excellent precision (195 ppm standard deviation out of transit) and no significant ramp systematics from charge trapping. For a smaller aperture size of 8 pixels (blue points), a larger jump (∼600 ppm) was observed in the middle of the primary transit, as well as a smaller one (∼100 ppm) toward egress, both due to tilts in a primary mirror segment which occurred during the observation (see Section 4.5.4 and Figure 11). The magnitude of the jumps can be reduced by increasing the extraction aperture size (orange points). Data are from PID 1442.

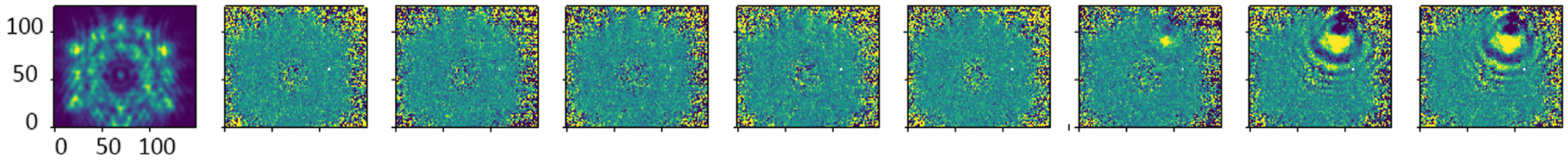

**Figure 11.** Weak lens data acquired at the same time as the light curve shown in Figure 10. The weak lens image at the start of the time series is shown along with ratios of the images leading up to the tilt event. The tilted segment is very apparent, and the time of the appearance of the tilt coincides with the jump in the white-light curve.

Table 3 compares the required, predicted, and on-orbit limiting point-source sensitivity of NIRCam imaging, using the flux calibration of 2022∼October. The limiting sensitivity is the flux density of the faintest point source that can be detected at signal to noise ratio $S/N = 10$ in an integration time of 10,000 s. For a representative wavelength for each of the short and long-wavelength channels, the table quotes the requirements values, the pre-launch predictions from the exposure time calculator, and the predicted performance assuming the measured on-orbit throughput, PSF, and detector noise levels. Table 3 shows that NIRCam imaging is substantially more sensitive than pre-launch expectations.

For the common case of background–limited broadband imaging of a point source, the integration time required (to reach a given S/N on a target of a given brightness) scales as the square of the sensitivity. As such, given Table 3, NIRCam deep imaging should proceed 3.3 and 2.4 times faster at 2.0 and 3.5 $\mu$m, respectively, than a system that just met requirements.

We quickly compare this limiting point source sensitivity to Hubble and Spitzer, again considering the faintest point source detectable at $S/N = 10$ in 10,000 s. For NIRCam at 1.5 $\mu$m, the sensitivity is 9 times better than WFC3-IR on Hubble.[56] For NIRCam at 3.5 $\mu$m, it is 68 times better than IRAC on Spitzer. Again, the integration time required to achieve a given limiting sensitivity, relative to these previous instruments, scales roughly as the square of these advantage factors for background-limited broadband imaging. Clearly, NIRCam imaging should detect faint objects substantially faster than pre-launch expectations.

[56] Comparing JWST/NIRCam F150W to HST/WFC3-IR F160W for a flat-spectrum source.

### *6.1.2. NIRCam Grism Time-series*

Observations of exoplanet HAT-P-14 b taken during commissioning (PID 1442) demonstrated that NIRCam grism time series spectroscopy is working well, meeting performance requirements (Schlawin et al. 2023, this issue) after only simple removal of systematic instrumental noise, known as detrending, by fitting an astrophysical lightcurve model times a polynomial and exponential model that are both functions of time. The noise for this mode is within 150% of the theoretical photon noise. Additional detrending and analysis will presumably move even closer to the photon noise limit.

With simple detrending, the standard deviation of the transit spectrum (as fit by a limb darkened transit model) was 91 ppm at $R = 100$, compared to the expected photon noise of 55 ppm. The settling time was observed to be 5–15 minutes. The throughput for NIRCam grism spectroscopy is 20%–40% higher than expected for most wavelengths. This may make

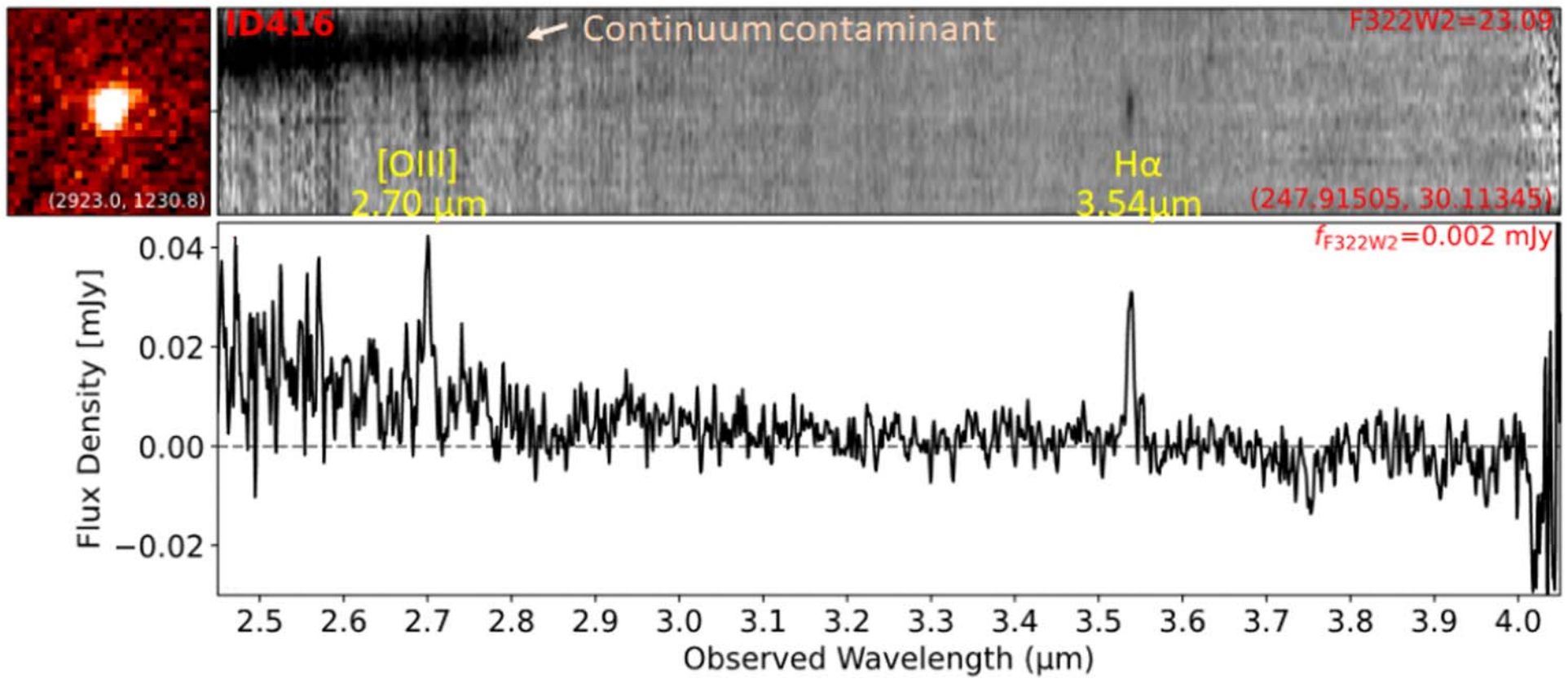

**Figure 12.** Spectrum of a $z = 4.39$ emission-line galaxy. This spectrum was detected serendipitously in 386 s of exposure time, in NIRCam wide field slitless spectroscopy mode data that targeted standard star P330-E for flux calibration, in PID 1076. Forbidden [O III] 5007 and H alpha are clearly detected.

saturation more likely for approved programs, particularly at long wavelengths.

Figures 10 and 11 illustrate the utility of monitoring the star in NIRCam's short wavelength channel with the weak lens, during such grism transit observations in the long wavelength channel. In this example, such monitoring with the weak lens captured two distinct tilts of primary mirror segments (see Section 4.5.4), which caused jumps in the grism data of 200–1000 ppm, depending on wavelength and extraction aperture size.

### 6.1.3. NIRCam Photometric Time-series

This mode is not heavily used in Cycle 1. In commissioning, the mode was checked out using a $J_{\mathrm{Vega}} = 14$ star with a short 500 s observation, in PID 1068. The performance was nominal: the standard deviation in the normalized flux was measured to be 0.62% (long wavelength channel) and 1% (short wavelength channel), both close to theoretical expectations.

### 6.1.4. NIRCam Wide Field Slitless Spectroscopy

NIRCam's Wide Field Slitless Spectroscopy (WFSS) has shown excellent performance through commissioning observations (see Rieke et al., this issue). With the F322W2 filter at 3.5 $\mu$m, in an integration of $10^4$ s, the continuum sensitivity (S/N=10 per resolution element) is 4 microJy, and the emission line sensitivity (line flux S/N = 10) is $2.8 \times 10^{-18}$ erg s$^{-1}$ cm$^{-2}$. Figure 12 illustrates the power of the NIRCam/WFSS mode, by showing a serendipitous detection of a line-emitting galaxy at $z = 4.39$ in the commissioning data. Note the strong detections of the [O III] 5007 Å and H$\alpha$ lines in the spectrum, which securely identify the redshift of this galaxy. A similar serendipitous detection of a $z = 6.11$ galaxy was reported by Sun et al. (2022).

### 6.1.5. NIRCam Coronagraphy

As described in Girard et al. (2022), the performance of the NIRCam coronagraphs exceed expectations. JWST commissioning demonstrated the 5$\sigma$ PSF-subtracted contrast of the 335R mask at 1″ to be roughly ten times better than requirements, achieving contrasts of $\sim 4 \times 10^{-5}$ to $\sim 4 \times 10^{-6}$. Coronagraphic target acquisition was demonstrated to be on par with requirements.

Stray light features have not been observed in the NIRCam coronagraph fields of view. At this time there is no evidence that stray light will impact coronagraphy or that mitigation strategies need to be taken.

## 6.2. NIRISS Performance

The Near Infrared Imager and Slitless Spectrograph (NIRISS) instrument is described in the companion PASP special issue paper Doyon et al. (2022). Here, we summarize the science performance of NIRISS as characterized during commissioning.

### 6.2.1. NIRISS Imaging (Parallel Only)

The imaging performance of NIRISS matches or exceeds expectations. Shortward of 2 $\mu$m, the measured throughput is about 20% higher than the instrument team's expectations, and about 25%–30% higher than the ETC's predictions. Longward of 2 $\mu$m, the throughput is about 5%–10% better than the instrument team's expectations, and 25% better than the prelaunch ETC predicted. Detector noise properties are similar to as measured on the ground, with a slightly higher "dark current" and total noise, likely due to residuals from uncorrected cosmic ray events. The sky background is lower than predicted before flight.

Table 4 compares the ETC-predicted and actual limiting point-source sensitivity of NIRISS imaging. The limiting sensitivity is the faintest point source that can be detected at

**Table 4**
NIRISS Limiting Point Source Sensitivity

| Wavelength ($\mu$m) | 1.15 | 2 | 3.5 | 4.4 |
|---|---|---|---|---|
| ETC prediction (nJy) | 13 | 10.2 | 14.5 | 22.8 |
| Actual (nJy) | 10.0 | 8.4 | 11.8 | 17.9 |

**Note.** What is quoted is the faintest flux density that can be detected for a point source at S/N = 10 in 10,000 s. Values are for wide-band filters. Smaller numbers are better. The requirement level was set at 13 nJy for the 3.5 $\mu$m filter.

signal to noise ratio S/N = 10 in an integration time of ten thousand seconds. The table quotes the pre-launch predictions from the exposure time calculator, and the predicted performance assuming the measured on-orbit throughput, PSF, and detector noise levels, and pre-flight background. This is predicted performance, not measured performance, as commissioning in general did not involve long integrations. Table 4 predicts that NIRISS parallel imaging should detect faint objects substantially faster than pre-launch expectations. The photometric stability is better than 1%, based on two measurements of the same standard star made 16 days apart.

The field distortion of NIRISS was calibrated using thousands of stars in the LMC astrometric field. Residual astrometric errors with respect to the catalog are 3 mas per axis. This is better than the requirement for accurate targeting of NIRISS sources with NIRSpec multi-object spectroscopy.

The NIRISS PSF is better or similar to pre-flight WebbPSF predictions as defined by encircled energy, FWHM, and ellipticity. There is very little field-dependence of the PSF as measured by these parameters. NIRISS imaging of very bright stars shows an extra diffraction spike offset from the vertical by between $-10^\circ$ and $+20^\circ$ with an angle that rotates smoothly as one moves from the right to the left edge of the detector. This spike is stronger at shorter wavelengths with an integrated intensity ~70% of the vertical diffraction spike for the shortest wavelength filter, F090W. The cause of this spike has been traced to diamond turning residual wave front errors in some of the NIRISS off-axis mirrors.

NIRISS images show a narrow band of excess stray light running almost horizontally across the detector, dubbed "the lightsaber." See Section 5.2.

Imaging ghosts were seen in ground testing of NIRISS and similar behavior is seen in flight. They are the result of internal reflections in the optical system. Ghost positions are predictable for each filter as they form at a position that is symmetric around the ghost axis point (GAP). The GAPs for each filter were measured during commissioning and the intensity of the ghosts was found to be ~1% of the original source intensity.

The cosmic ray rate at L2 is similar to predictions. However there is a much higher rate of large "snowballs" that appear as diffuse, mostly circular, events that usually saturate in their centers. See Section 6.6.

### *6.2.2. NIRISS Single Object Slitless Spectroscopy*

Time series observations of a spectrophotometric standard A-star (BD+601753, $K_{s,\mathrm{Vega}} = 9.6$, PID 1091, duration 5 hr) were used for flux calibration during commissioning. The median precision obtained in order 1 was 147 ppm and order 2 was 206 ppm, binned to a 22 s integration time. This provides an independent test of the stability and precision achieved for a non-variable star, with only a low-level polynomial trend observed in spectral order 1 and no significant trends observed in orders 2 and 3. This observation was affected by a tilt event approximately in the middle of the time series resulting in a flux jump of a few 100 s ppm. The tilt event was easily detected by monitoring the PSF shape along the spatial direction, more specifically by measuring the second derivative of the PSF which is a good proxy of the FWHM. The flux jump was demonstrated to be achromatic.

Observations of the exoplanet HAT-P-14 b (PID 1541, duration 6 hr) returned a point-per-point median precision in the transit depth of 85 ppm at $R = 100$ for order 1, and 90 ppm at $R = 100$ for order 2. The weighted scatter of the spectrum itself was 92 ppm for order 1 and 85 ppm for order 2. Errors in the transit depth are within <10%–20% from expectations at this resolution. A tilt event was also noted early in the sequence well before ingress. The NIRISS SOSS mode readily meets performance requirements.

Throughput in this mode is 25% better for order 1 near the blaze wavelength at 1.3 $\mu$m, and ~50% better for order 2. There appears to be no significant noise penalty to operate at 70%–75% of the saturation level. Observations should generally not exceed 35,000 ADUs (56,000 e-) to avoid extra noise. There is also a trade-off with observing efficiency; users may opt to saturate part of the spectrum to improve the duty cycle.

### *6.2.3. NIRISS Wide Field Slitless Spectroscopy*

The throughput for NIRISS wide field slitless spectroscopy (WFSS, Willott et al. 2022; Figure 13) is generally 30% better than expected from the prelaunch ETC. As a result, the predicted on-sky sensitivities (assuming the pre-flight background model) are better than the prelaunch ETC predictions by 7%–20%, similar to the results in imaging mode at these wavelengths. The GR150C filter has higher throughput than GR150R at wavelengths below 1.2 $\mu$m, likely due to a different anti-reflective coating, so is preferred for F090W and F115W if only one grism is used (normally both are used to mitigate contamination).

Trace positions, curvature, dispersion and spectral resolution for WFSS are close to those measured on the ground. In addition to dispersed versions of imaging ghosts, additional ghosts due to the grisms are seen, but at a relatively low intensity level. The lightsaber scattered light feature is also apparent for WFSS. For the typical intensity where the lightsaber is dominated by zodiacal light, this emission is included in the WFSS background reference files, so will be subtracted off in the pipeline.

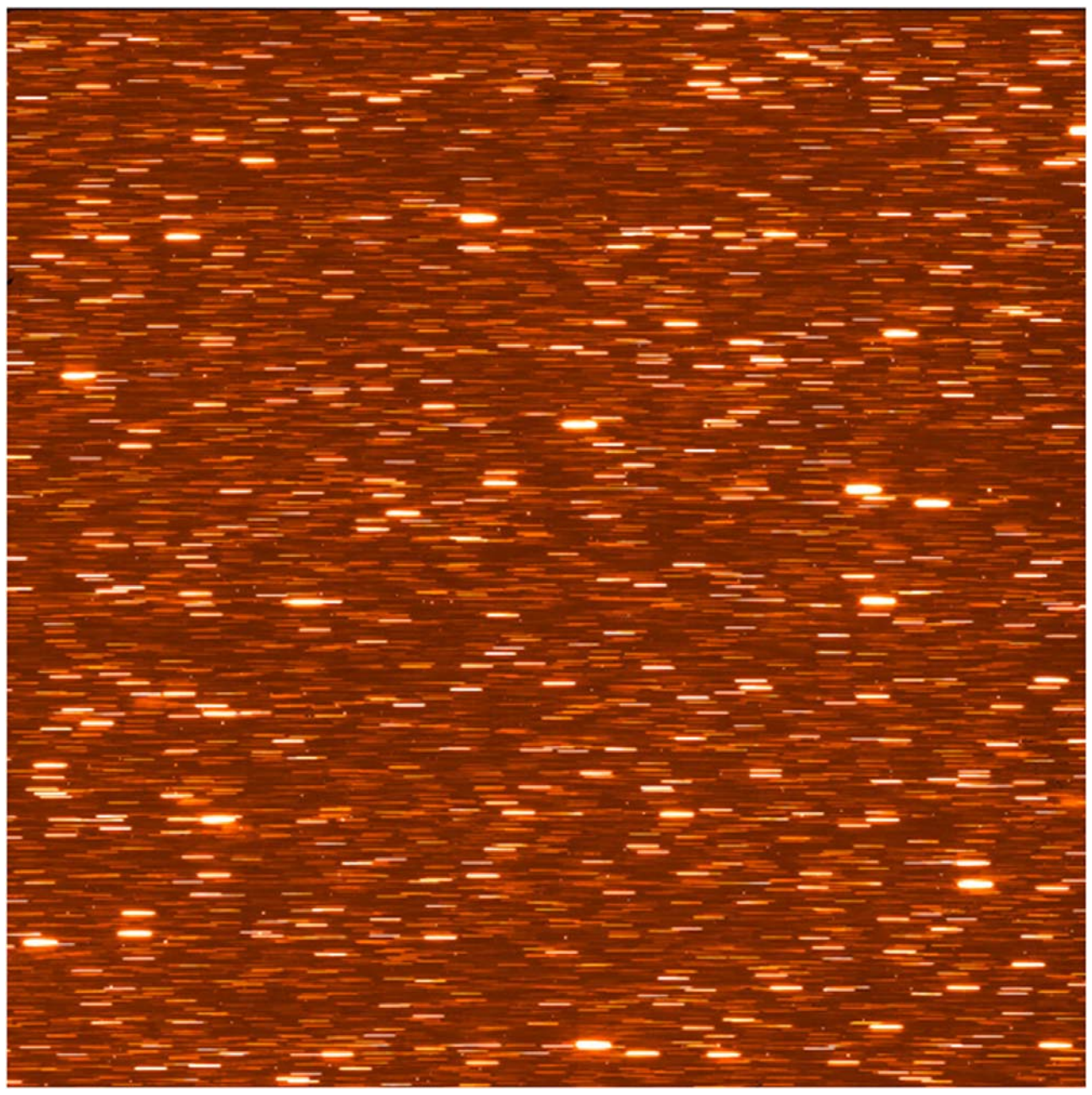

**Figure 13.** NIRISS WFSS. Simultaneous spectroscopy of thousands of stars in the NIRISS focus field of the LMC. Configuration is grism GR150C and filter F115W. The field of view of the detector is 2.′2 on a side. Data from PID 1085.

### *6.2.4. NIRISS Aperture Masking Interferometry*

The AMI mode (Sivaramakrishnan et al. 2022) features a seven-hole non-redundant mask that enables high-contrast ($10^{-3}$–$10^{-4}$) imaging at sub $\lambda/D$ angular separations (0.″1–0.″5) over three medium-band filters (F380M, F430M, F480M). This mode was successfully demonstrated through the easy detection of AB Dor C, a companion with a separation of ∼0.″3 with a contrast ratio of 4.5 mag. The noise floor of this data set is 6.5–7.0 mag (3$\sigma$), very close to the photon noise floor limit of ∼7.5 mag. This is the first space-based demonstration of both infrared interferometry and non-redundant aperture masking. The nominal operational concept for AMI requires staring (rather than dithering) on the science target, followed by a similar observation on an isolated reference star. Target acquisition (TA) places targets at the same detector location, and TA accuracy well within 0.1 pixel was demonstrated. Kernel Phase Interferometry (KPI) is the full pupil generalization of AMI used without the NRM mask but using a similar Fourier-based removal of instrument effects from the data (Kammerer et al. 2023). Like AMI, KPI also enables sub $\lambda/D$ imaging but with better throughput at the price of a lower contrast at small separation. Observations were performed on 4 targets, all presumed to be single stars, but one was found to have a companion at 0.″15 with a contrast of 1.7 mag. Interferometry with JWST has shown unsurpassed fringe amplitude stability, promising valuable complementarity to ground-based interferometry's significantly higher resolution.

### *6.3. NIRSpec Performance*

A detailed description of the Near-Infrared Spectrograph (NIRSpec) instrument and of its pre-launch performance can be found in a series of four papers: overview (Jakobsen et al. 2022); multi-object spectroscopy (MOS) mode (Ferruit et al. 2022); integral field spectroscopy (IFS) mode (Böker et al. 2022); and exoplanet time series (Birkmann et al. 2022). The on-orbit science performance of NIRSpec is described in the companion PASP special issue paper Böker et al. (2023). Here we briefly summarize those results.

All modes of NIRSpec are working well, in general better than pre-launch expectations. Both of NIRSpec's two detectors show noise levels, in the actual cosmic ray environment of L2, similar to or lower than ETC predictions. The excellent optical quality of the telescope translates to lower-than-expected slit losses in the multi-object and fixed slit modes of NIRSpec. For the 200 mas wide microshutters, this translates to increased photon conversion efficiency of 2.5% at 5 $\mu$m, >7.5% below 3 $\mu$m, and >10% below 1 $\mu$m. NIRSpec bright object time series mode has demonstrated precision in the transit depth of 50–60 ppm per point (Espinoza et al. 2023). Both target acquisition methods that are specific to NIRSpec, wide-aperture target acquisition (WATA) and microshutter assembly target acquisition (MSATA), are working well.

Figure 4 of Böker et al. (2022) shows the measured on-orbit sensitivity of NIRSpec for both multi-object spectroscopy and integral field spectroscopy. Across the board, the sensitivity is better than pre-launch predictions.

The operability rate of the un-vignetted microshutters is 82.5% (Rawle et al. 2022), with electrical short masking now the primary cause of non-operable shutters. The resulting multiplexing levels are within 10% of the pre-launch levels and are still excellent, with the possibility, for high target densities, to observe more than 200 scientific targets at a time at low spectral resolution or close to 60 at medium / high spectral resolution (see Figure 14).

### *6.4. MIRI Performance*

The Mid-Infrared Instrument (MIRI; Wright et al. 2023, this issue) is the only instrument on JWST that operates beyond 5 $\mu$m; as such it supports a broad range of measurement types: (1) standard imaging; (2) high contrast (coronagraphic) imaging; (3) low resolution spectroscopy (LRS) with and without a slit; and (4) medium resolution integral field unit spectroscopy (MRS).

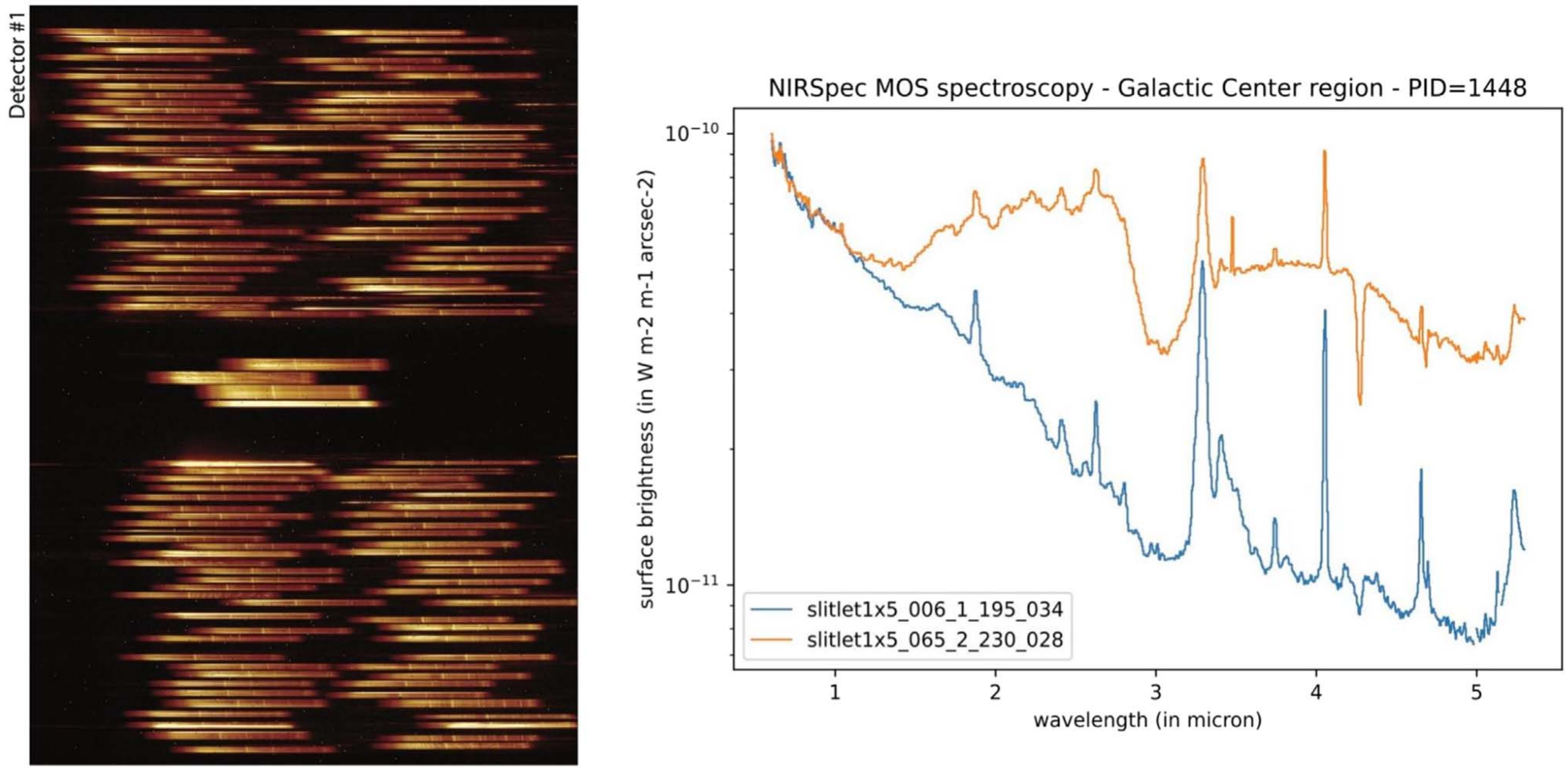

**Figure 14.** Multi-object spectroscopy with the NIRSpec microshutter array. The low-spectral resolution prism mode was used, with a total of 235 microshutter slitlets opened, to capture spectral features from the diffuse interstellar medium in a region close to the Galactic Center. Extracted example spectra (not flux calibrated) show multiple emission and absorption line features. Data from PID 1448.

The MIRI imager uses most of a 1024 × 1024 pixel detector array to provide a field of view of 74″ × 113″ with eight broadband filters, providing bands starting at 5.6 $\mu$m and spaced every factor of ~1.2–1.3 to 25.5 $\mu$m. An additional 6% bandwidth filter is centered on the aromatic feature at 11.3 $\mu$m. The images are all diffraction limited, and they are Nyquist sampled for wavelengths longer than 6.25 $\mu$m. The overall throughput of the OTE and MIRI imager exceeds the prelaunch expectations, particularly for wavelengths at 10 $\mu$m and longer. The result is that the imager sensitivity is improved over prelaunch predictions.

JWST provides subarcsec imaging in the mid-infrared, with a beam 50 times smaller in area than that of the Spitzer Space Telescope, the previous most capable space infrared telescope. Figure 15 illustrates this capability, which opens an entirely new realm of study of the structure of mid-infrared sources. Where MIRI is limited by natural backgrounds (for wavelengths of 5 to at least 12.5 $\mu$m, Rigby et al. 2023, this issue), its sensitivity for point sources is about 50 times better than that of the Spitzer Space Telescope. This gain is reduced at longer wavelengths due to telescope emission, but in deep exposures the lack of confusion noise with the small beam of MIRI still provides significant gains over Spitzer.

Four coronagraphs lie along a side of the imaging field of view, optimized for wavelengths of 10.58, 11.30, 15.50, and 23 $\mu$m, and providing raw contrasts of $\sim 10^4$ (at $6\lambda/D$). The MIRI coronagraphs are performing significantly better than anticipated (Boccaletti et al. 2022), in part due to the excellent image quality of the telescope. Other key factors are the achieved precision of alignment of MIRI and the ISIM to the telescope in pupil shear and focus, both of which are better than the budget allocations. In common with other coronagraphs on JWST, subtraction of a PSF reference star is necessary to achieve the best performance. This technique is very sensitive to high order wave front error (i.e., the shape of the PSF wings) and thus scheduling needs to take account of tilt events and routine telescope mirror alignments.

The MIRI low resolution spectrometer (LRS) also lies to one side of the imager. It has very high throughput (~80%) and nominal resolution of $\lambda/\Delta\lambda = 100$, designed for spectroscopy of very faint objects. Its performance is optimized for 5–10 $\mu$m, but it is usable to 14 $\mu$m. There is also a slitless mode of LRS for exoplanet transits. During commissioning, a transit spectrum of the exoplanet L168-9b was obtained, demonstrating calibration to ~25 ppm with a spectral resolution of ~50 at 7.5 $\mu$m (Bouwman et al. 2023). LRS transit spectroscopy will be used to study organics and water in the atmospheres of exoplanets, helping determine abundances of these molecules and whether non-equilibrium chemistry is at work influencing them.

The MIRI medium resolution spectrometer (MRS) covers by design the full 5–28.3 $\mu$m range (currently calibrated to 27.9 $\mu$m). It uses integral field units as its inputs, with fields increasing with increasing wavelength from $3\overset{''}{.}2 \times 3\overset{''}{.}7$ to $6\overset{''}{.}6 \times 7\overset{''}{.}7$. The spectra and spatial information are arranged over two 1024 × 1024 pixel detector arrays to provide spectral resolution, $\lambda/\Delta\lambda$, of $\gtrsim$ 3000 for wavelengths shorter than $\sim$11.7 $\mu$m, and $\gtrsim$1500 out to 28.3 $\mu$m, along with near-diffraction-limited imaging. The high spectral resolution, sensitivity, and imaging capability of the MRS together provide breakthrough capabilities for mid-infrared spectroscopy. For example, MRS gives access to many molecular transitions (e.g., organic molecules, H2) and to the full suite of neon fine structure lines: [Ne II] 12.81 $\mu$m, [Ne III] 15.56 $\mu$m, [Ne V] 14.32 $\mu$m, and [Ne VI] 7.64 $\mu$m, which are very powerful for determining the excitation mechanism of emission line objects. Figure 16 illustrates the use of these capabilities to dissect the planetary nebula NGC 6543. These data also illustrate well the quality of the calibration of the MRS distortion, fields of view and astrometry achieved during commissioning. While the point source illumination used during ISIM testing on the ground was too faint at mid-IR wavelengths to characterize well these aspects of the MRS before launch, the commissioning data confirm the design expectations.

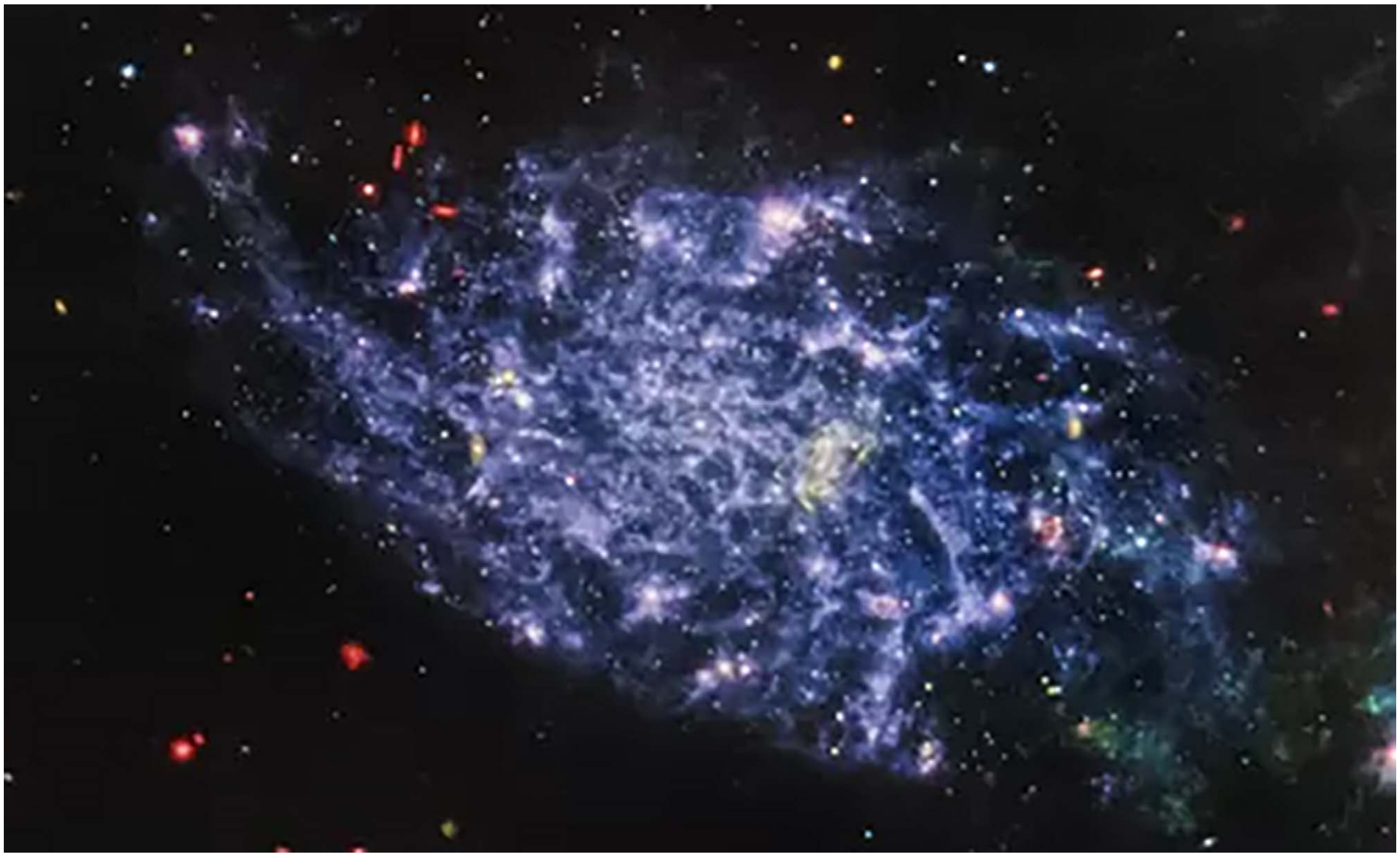

**Figure 15.** NGC 7320, the foreground galaxy in Stephan's Quintet, as seen in MIRI imaging. The major axis of the galaxy is about 2′ long. The MIRI image is in three bands: F770W, F1000W, and F1500W, respectively shown in false color as blue, green, and red. In the image, red denotes dusty, star-forming regions, and blue can show either stars, or the strong and prominent 7.7 $\mu$m aromatic band (which dominates the image). What appears in the visible to be a typical dusty spiral galaxy is lit up in this MIRI image, with complex structure tracing where aromatic molecules are heated by hot stars. Data from PID 2732 (Pontoppidan et al. 2022). Credit: NASA/ESA/CSA/STScI.

### 6.5. Using Two Science Instruments in Parallel

JWST supports parallel observing, in which two science instruments are used simultaneously. There are two types of parallel: coordinated parallels and pure parallels.

#### 6.5.1. Coordinated Parallels

Coordinated parallels are when two science instruments are used together for the same observing program. A total of 170 coordinated parallel visits were successfully executed during science instrument commissioning (Figure 17), and worked well with only one issue (discussed just below). Parallel operations are designed such that mechanism movements of each instrument do not disturb the operation of the other; this coordination is working as designed.

While there was no requirement that all of the eleven coordinated templates be exercised during commissioning, in the end seven of the templates were. The four templates that were not exercised are: NIRCam Imaging + NIRISS WFSS;

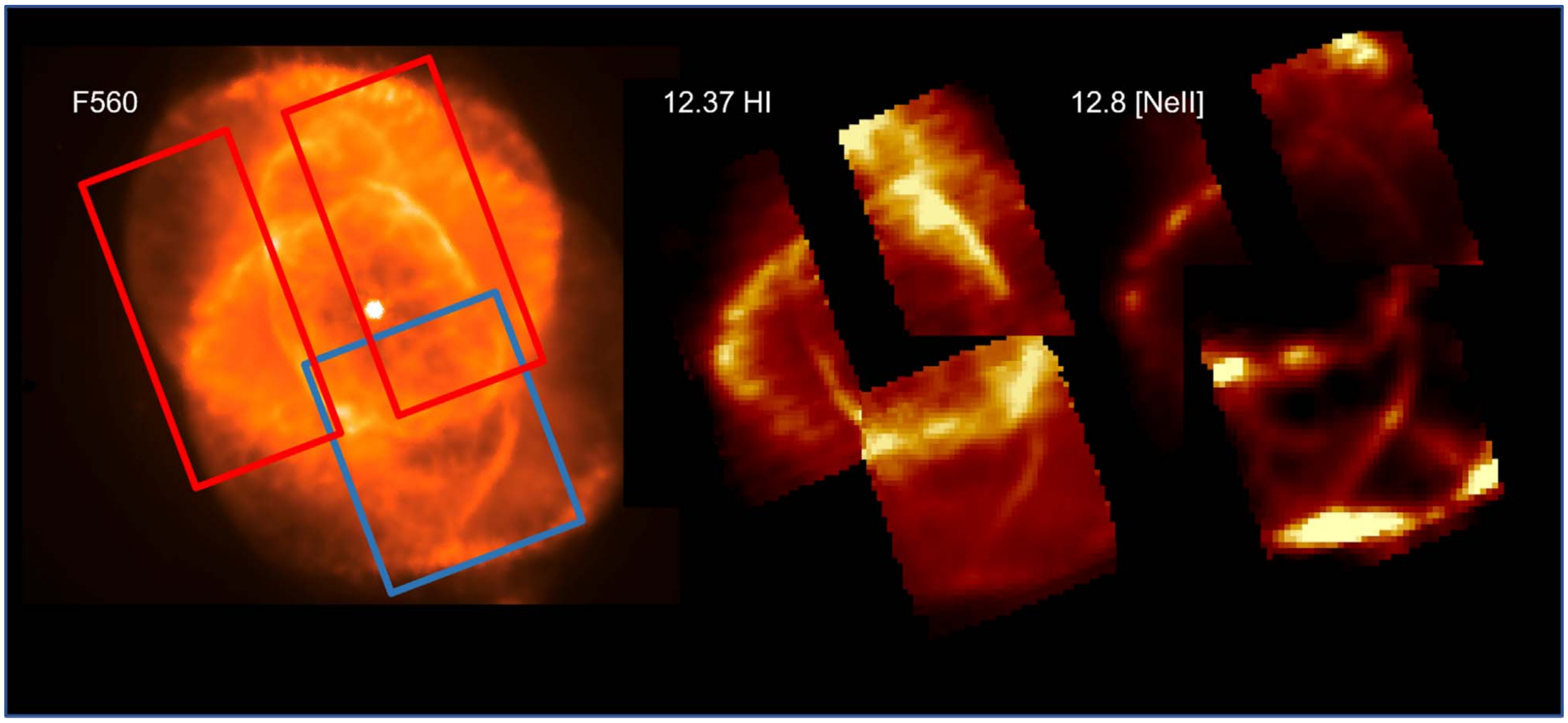

**Figure 16.** NGC 6543, the Catseye planetary nebula, imaged at 5.6 $\mu$m (left panel), and dissected by the MIRI MRS in mosaics using its integral field unit (right two panels). The maximum extent of the nebula in the image is ∼25.” The two layers in the spectral/spatial cube show the nebula at sub-arsecond resolution in Humphreys $\alpha$ 12.37 $\mu$m and the fine structure line of [Ne II] 12.8 $\mu$m. MRS has four integral field units; the field of view increases with wavelength. Data are from programs PID 1023, 1031, and 1047. Credit: B. Vandenbussche

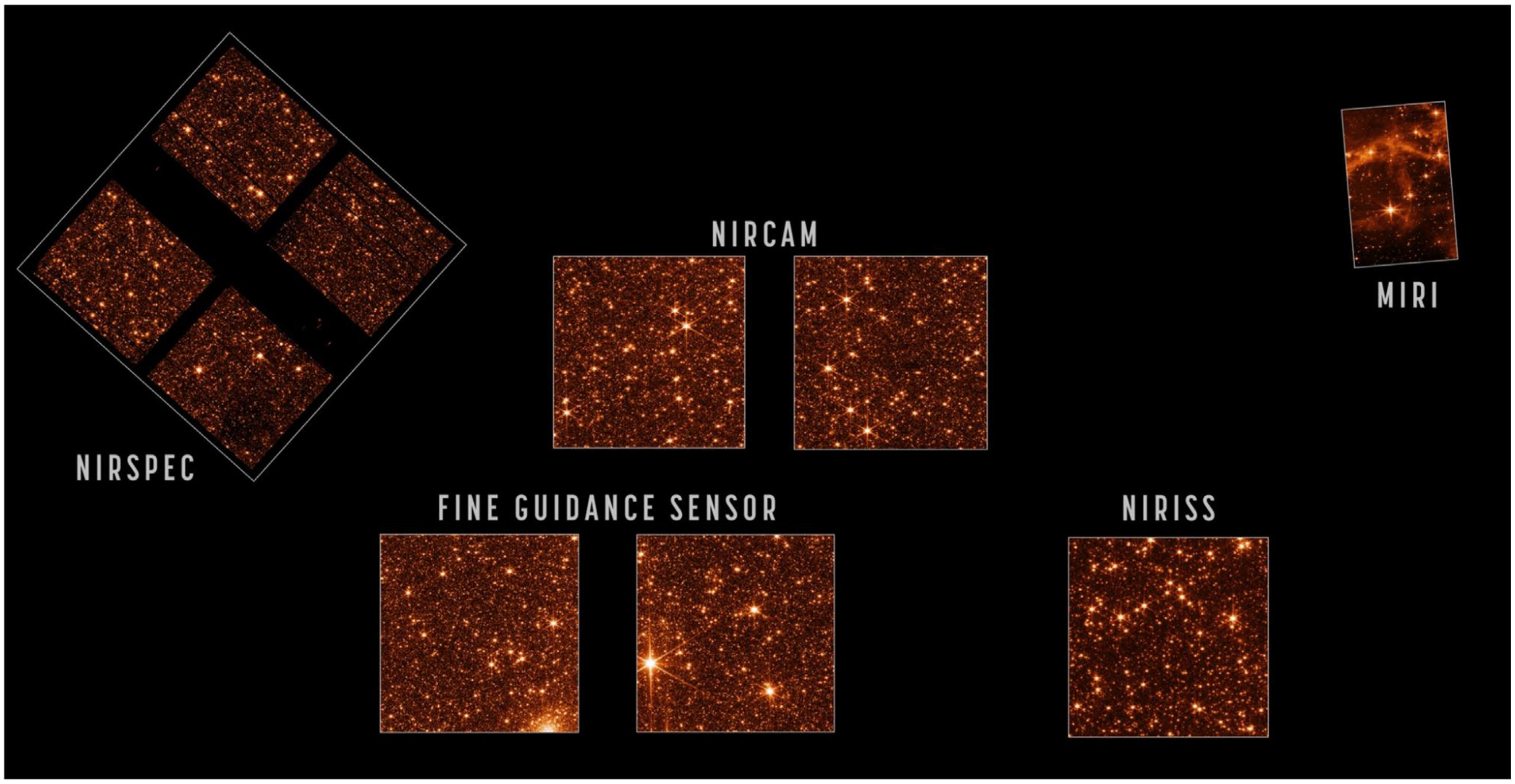

**Figure 17.** Observations of the JWST astrometric calibration field and surroundings in the Large Magellanic Cloud. These observations from PID 1473, previously released, were taken just after the completion of multi-instrument optical alignment to assess and demonstrate image quality in all instruments. This also served as an engineering test of coordinated parallel operations: this included the first uses of NIRCam+MIRI imaging, MIRI+NIRCam imaging, and NIRSpec MOS spectroscopy +NIRCam imaging, as well as NIRCam+NIRISS imaging and NIRCam+FGS imaging, with many combinations of filters and parallel-optimized dither patterns. The pointings on sky are all shown in the correct relative orientations and scales relative to one another; the figure is approximately 18′ horizontal by 9′ tall.

NIRCam WFSS + MIRI Imaging; NIRCam WFSS + NIRISS Imaging; and MIRI Imaging + NIRISS WFSS. We see no reason why these templates should not work, as these WFSS templates are identical to the imaging templates except for putting a grism in the beam instead of a filter.

The one issue with parallel observing that was identified was this: when all 10 NIRCam detectors were used in parallel with another instrument, data from one instrument could under some circumstances be partially overwritten by data from the other instrument. Flight software was patched on 2022 June 24 to fix this issue. Data taken earlier may be affected by partial data overwrite.

#### *6.5.2. Pure Parallels*

Pure parallels are when an observing program makes use of parallel observing opportunities from other accepted proposals. Thus, in pure parallel mode two science programs execute at once. Pure parallel mode was not exercised in commissioning. Pure parallel programs were not scheduled for the first four months of Cycle 1; they were enabled in 2022 November. The concern had been the tight (11%) margin on data downlink in Cycle 1. Data volume is used as a constraint when scheduling JWST observations; the schedule is built to keep the onboard solid state recorder from filling, since if the recorder fills JWST would halt observing and sit idle until the next ground contact. The project will continue to assess how well data volume is managed through scheduling and downlink performance during science operations.

### *6.6. Cross-instrument Detector Topic: Cosmic Rays*

Observed cosmic ray rates and properties are largely in line with expectations. The vast majority of cosmic ray impacts directly affect only one or 2 pixels, but there are also uncommon events that affect hundreds of pixels. These large events are colloquially termed snowballs. There are also large radiation events that affect the MIRI detectors, which are called "shower" events.[57]

The current JWST data reduction pipeline handles the first order effects from cosmic ray events but the large number of electrons that result from a snowball event have secondary effects that are not currently corrected in the pipeline. Residuals have a circular appearance with alternating light and dark bands a few tens of pixels across. Dithering exposures is the current recommended mitigation strategy. A four-point or larger dither pattern will allow the pipeline outlier detection routine to significantly improve the final combined image. Work is in progress to improve the calibration pipeline's detection and handling of regions affected by snowballs.

[57] https://jwst-docs.stsci.edu/jwst-mid-infrared-instrument/miri-features-and-caveats

## 7. Science Operations Status

Science operations includes managing the proposal process, preparing visits for execution, processing data from the observatory, and making data products available via the MAST archive. JWST science operations at STScI provided outstanding service during commissioning and the transition to Cycle 1. Complex software and processes worked thanks to many tests and rehearsals prior to launch. Planned updates and unexpected issues were handled via operational work-arounds and software updates. Nevertheless, some significant issues remain and warrant attention, as discussed below.

The Operations Scripts Subsystem (OSS) carries out the science observations by executing the Observation Plan (OP), which is uploaded weekly and executed by OSS autonomously, often while JWST is not in contact with the Mission Operations Center. OSS was used to conduct most of the telescope and science instrument commissioning activities, and has been patched several times to resolve issues identified during commissioning.

During commissioning, the JWST Exposure Time Calculator (ETC) reflected pre-launch expectations. As described above, end-to-end flight performance is typically better, in some cases by a significant amount. The Cycle 2 Call for Proposals, released on 2022 November 15, uses sensitivities measured with in-flight data. In the beginning of Cycle 1, observers were able to use performance information above and in JDox to scale ETC results. For the typical case of higher than expected throughput, the main concern for Cycle 1 programs was unexpected saturation during target acquisition or early in an integration. Saturation later in an integration will generally yield S/N comparable to or better than predicted by the ETC.

The Astronomer's Proposal Tool (APT) and the Micro-shutter Array Planning Tool now reflect flight measurements of aperture placement in the focal plane and distortion across instrument apertures. Additional refinements are expected, but the impact on observers should be negligible. Functionality will also evolve slightly to accommodate insights from commissioning.

Long range plan windows have been published for most Cycle 1 visits. Investigators can search for program information and click the Visit Status Information link for more information. Whenever possible, visits are scheduled and executed within their assigned plan window, but operational issues may cause plan windows to change, visits not to schedule, or scheduled visits not to execute. This is particularly true early in Cycle 1. Visits that do not execute are usually rescheduled.

JWST uses the Guide Star Catalog (GSC) to specify astrometry and photometry of guide stars and reference stars for guide star acquisition. During commissioning, a few guide star acquisitions failed because the guide "star" was actually a galaxy resolved by FGS or because coordinates in the GSC were wrong by an arcsec or more. The same will happen during

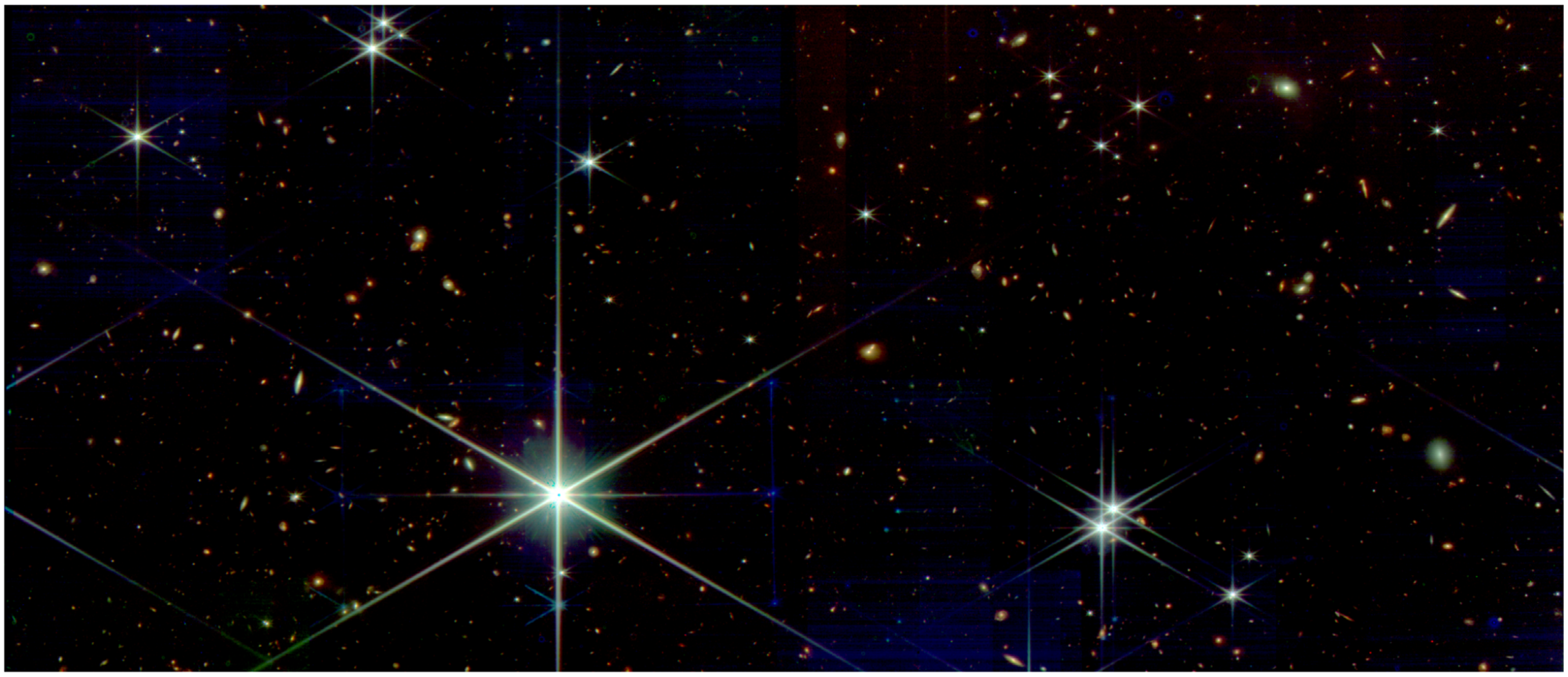

**Figure 18.** Multi-wavelength image mosaic of the Fine Phasing field, demonstrating the efficacy of automated data processing. This is a color version of the image that was released in 2022 March just after the completion of telescope fine phasing; the blue, green and red channels show NIRCam F115W, F200W, and F356W. It was produced simply by retrieving the automatically produced Level 3 mosaic data products from MAST, and opening them in ds9 to make an RGB image. The only manual step applied was a simple median subtraction of background level, used primarily to remove a temporarily high background in the 3.5 $\mu$m channel in these early data, which appeared because the instruments were not yet fully cooled. The automated pipeline generates a nearly science-ready product with all mosaic tiles stitched and the several filters registered together. The field of view is $6'$ by $2\rlap{.}'6$. Data from PID 1160 observation 22.

Cycle 1, but the frequency of failures will decrease with improvements to the GSC and operational procedures.

After processing all data taken during commissioning, the JWST data management subsystem generated 1.5M files, requiring 55 TB to store a single copy. Users should be selective about which types of data products they download to conserve network bandwidth and storage. Users can download and rerun the JWST calibration pipeline with custom parameters tailored to their needs, or even modify or add steps to the calibration pipeline.

JWST data are now available in the MAST archive (see Accessing JWST Data), as well as in the JWST archives at ESA and CSA. See Figure 18.

Because Cycle 1 observing began shortly after observatory commissioning, data products at first had relatively poor calibration accuracy and contained calibration artifacts. As instrument teams generate new calibration reference files throughout Cycle 1, the calibration accuracy will improve. Science operations will reprocess old data and make improved data products available in MAST, as needed when new calibration files become available.

Users should be especially vigilant about astrometric and photometric calibration errors in data products downloaded early in Cycle 1. Errors in guide star coordinates or focal plane geometry will propagate to the world coordinate system (WCS) for exposure-level data products. Higher-level products (e.g., mosaics) may use tweakreg and Gaia sources in each image to transform the WCS onto the Gaia frame. Photometric calibration will reflect pre-launch expectations until calibration reference data are updated.

Users should note the calibration pedigree of their data files and of data files used in publications. Consult JWST documentation for a description of known calibration issues and their resolution in successive versions of calibration reference data and software.

Data processing typically takes about a day, but can take longer if downlink, data transfer, or processing issues arise. For complicated modes (e.g., multi-object spectroscopy) or associations containing many exposures, the calibration pipeline can take several hours to run or even days in extreme cases. Performance will improve over time, but for now the main focus is functionality.

### *7.1. JWST User Documentation System (JDox) Updates*

The JWST User Documentation system (JDox) provides comprehensive information about the JWST Proposing process, the Observatory and science instruments, science data characteristics and data access, data pipeline processing and calibration, as well as introductions to post pipeline data analysis tools and training materials.

JDox was updated on 2022 July 12 to describe data features and image artifacts seen in flight data, known shortcomings in current data pipeline products, and other articles to help users understand the status of the Observatory and science instruments. The next major release of JDox coincided with the release of the Cycle 2 Call for Proposals on 2022 November 15.

Users can refer to the Latest Update information box at the bottom of each JDox article to see when that article was updated. JDox also maintains a summary page showing recently changed articles by titles and when they were updated.

## 8. Conclusions

This article summarizes the science performance of JWST as characterized by the six month commissioning period. Almost across the board, the science performance of JWST is better than expected. The optics are better aligned, the point-spread function is sharper with higher encircled energy, and the optical performance is more time-stable than requirements. The fine guidance system points the observatory several times more accurately and precisely than required. The mirrors are cleaner than requirements, which translates into lower-than-expected levels of near-infrared stray light, meaning that the $<5\ \mu$m sky background will be darker for JWST than expected. As it was designed to be, JWST is indeed limited by irreducible astronomical backgrounds, not by stray light or its own self-emission, for all wavelengths $<12.5$ micron. The science instruments have generally higher total system throughput than pre-launch expectations. Detector noise properties are similar to ground tests, albeit with higher rates of cosmic rays, as expected in deep space. Collectively, these factors translate into substantially better sensitivity for most instrument modes than was assumed in the exposure time calculator for Cycle 1 observation planning, in many cases by tens of percent. In most cases, JWST will go deeper faster than expected. As a key example, NIRCam deep imaging of point sources to a given depth should proceed 3.3 and 2.4 times faster at 2.0 and 3.5 $\mu$m, respectively, than a system that just met requirements. In addition, JWST has enough propellant onboard to last at least 20 yr.

This characterization of science performance undergirds the key conclusion of commissioning: that JWST is fully capable of achieving the discoveries for which it was built. JWST was envisioned "to enable fundamental breakthroughs in our understanding of the formation and evolution of galaxies, stars, and planetary systems" (Gardner et al. 2006)—we now know with certainty that it will. The telescope and instrument suite have demonstrated the sensitivity, stability, image quality, and spectral range that are necessary to transform our understanding of the cosmos through observations spanning from near-earth asteroids to the most distant galaxies.

Commissioning proved the observatory's capabilities through approximately 2300 visits of commissioning observations, which exercised the same science instrument modes that will be used in normal science operations. All four science instruments have demonstrated the ability to precisely capture spectra of transiting exoplanets with initial precision better than 100 ppm per measurement point, with limiting performance expected to be well below that level.[58] JWST has tracked solar system objects at speeds up to 67 mas $s^{-1}$, more than twice as fast as the requirement. JWST has detected faint galaxies with fluxes of several nano-Jansky, and observed targets as bright as Jupiter. JWST has obtained infrared spectra of hundreds of stars simultaneously in a dense starfield toward the Galactic center, as well as integral field spectroscopy of planetary nebulae and a Seyfert nucleus at unprecedented sensitivity. Data from these and all other commissioning activities are available to the scientific community via the MAST archive.

As each of JWST's seventeen science instrument modes finished its commissioning activities, it was reviewed against mode-specific readiness criteria for science instrument performance. All JWST observing modes have been reviewed and confirmed to be ready for science use. In most cases, the modes surpass performance requirements.

Continued analyses as well as Cycle 1 calibrations will further improve the characterization of the science instruments. Updated knowledge has been reflected in updates to data pipeline reference files and algorithms, as well as thorough revisions to the JDox documentation and updates to the JWST proposal tools that were timed to support the Cycle 2 Call for Proposals.[59]

JWST is the product of the efforts of approximately 20,000 people in an international team. Commissioning JWST and characterizing science performance is the result of tremendous effort by the JWST commissioning team over six months. The achieved performance is the result of efforts over the many years leading to launch by team members across much of the globe. Given the measured performance described in this document, the JWST mission entered Cycle 1 having demonstrated that the observatory exceeds its demanding pre-launch performance expectations. With revolutionary capabilities, JWST has begun the first of many years of scientific discovery.

This work is based on observations made with the NASA/ESA/CSA James Webb Space Telescope. The data were obtained from the Mikulski Archive for Space Telescopes at the Space Telescope Science Institute, which is operated by the Association of Universities for Research in Astronomy, Inc., under NASA contract NAS 5-03127 for JWST. These observations are associated with many programs from the commissioning period. Technical contributions were carried

[58] That time series precision of $< 100$ ppm is measured at spectral resolutions of $R = 100$ at wavelengths below 5 $\mu$m and $R = 50$ at longer wavelengths, with minimal detrending.

[59] The Cycle 2 Call for Proposals was released on 2022 November 15, with proposals due 2023 January 27.

*Facilities:* JWST (NIRCam, NIRISS, NIRSpec, MIRI).

*Software:* Pandeia (Pontoppidan et al. 2016).

## ORCID iDs

Jane Rigby https://orcid.org/0000-0002-7627-6551
Michael McElwain https://orcid.org/0000-0003-0241-8956
Pierre Ferruit https://orcid.org/0000-0001-8895-0606
Marcia Rieke https://orcid.org/0000-0002-7893-6170
Chris Willott https://orcid.org/0000-0002-4201-7367